\RequirePackage{snapshot}
\documentclass[hyperref,letterpaper,twocolumn,pre,floats,aps,superscriptaddress,amsmath,amssymb,nofootinbib]{revtex4-1}

\usepackage{mathptmx}

\usepackage{graphicx}
\usepackage{multirow}
\DeclareGraphicsExtensions{.pdf,.eps,.png,.jpg} 
\usepackage{bm}
\usepackage{amssymb,latexsym,amsmath,amsfonts}     
\usepackage{hyperref}
\usepackage{epstopdf}

\makeatletter
\newcommand*{\centerfloat}{
  \parindent \z@
  \leftskip \z@ \@plus 1fil \@minus \textwidth
  \rightskip\leftskip
  \parfillskip \z@skip}
\makeatother

\usepackage{xcolor}
\usepackage{comment}

\usepackage[normalem]{ulem}
\renewcommand{\BibitemShut}[1]{}

\graphicspath{{./figs/}}  

\begin{document}

\title{Crossover Behavior in the Packing and Assembly of
  Multivalent Lock-and-Key Colloids}
\author{N. Khalid \surname{Ahmed}}
\affiliation{Department of Chemical Engineering,
University of Michigan, Ann Arbor, MI 48109-2136, USA}
\author{Greg \surname{van Anders}}
\affiliation{Department of Chemical Engineering,
University of Michigan, Ann Arbor, MI 48109-2136, USA}
\author{Elizabeth R. \surname{Chen}}
\affiliation{Department of Chemical Engineering,
University of Michigan, Ann Arbor, MI 48109-2136, USA}
\affiliation{School of Engineering and Applied Sciences,
Harvard University, Cambridge, Massachusetts 02138, USA}
\author{Sharon C. \surname{Glotzer}}
\affiliation{Department of Chemical Engineering,
University of Michigan, Ann Arbor, MI 48109-2136, USA}
\affiliation{Department of Materials Science and Engineering,
University of Michigan, Ann Arbor, MI 48109-2136, USA}
\date{\today}

\begin{abstract}
  Emergent behaviors occur in a vast array of systems across many scales, and
  are of fundamental physical importance because of the intrinsic difficulty in
  linking microscopic system properties to macroscopic behaviors. Here we study
  the emergent self-assembly behavior of model systems of recently synthesized
  families of concave dimpled hard spheres, or lock-and-key
  colloids. We find that as dimple size increases each family exhibits a crossover
  from a structure that does not reflect the particle symmetry to one that does
  and, surprisingly, the point at which this crossover occurs is approximately
  independent of the particle symmetry. Using a combination of numerical and
  analytic techniques we study systems at infinite and finite pressure, and find
  different common control parameters in each limit. Our results suggest there
  exists a set of experimentally realizable colloidal systems that exhibit
  complex emergent behaviors that can be traced to a common
  underlying microscopic control parameter.
\end{abstract}

\maketitle

\section{\label{sec:intro}Introduction}
Physical systems from flocking animals~\cite{Cucker2007a} to heavy fermion
materials~\cite{Aynajian2012} exhibit emergent macroscopic behaviors that are
intrinsically difficult to predict from their microscopic
properties~\cite{Anderson1972}. Colloidal systems are fertile ground for
investigating emergent behaviors for three reasons: a wide range of colloidal
systems manifest emergent behaviors because they have entropy-driven phase
behavior~\cite{Kamien2007,entint,escoengent}; there is a broad range of
techniques for manipulating colloids
experimentally~\cite{pawarkretzchmar,Sacanna2011,Shah2014a}; and colloidal
systems can be investigated with a variety of techniques including \textit{in
situ} confocal microscopy~\cite{Prasad2007}. Here, we study how microscopic
properties affect the emergent macroscopic behavior of a model family of
anisotropically shaped colloidal particles.

\begin{figure}
  \begin{tabular}{cccccccccc}
   \multicolumn{1}{l}{\textbf{(a)}} & & & & & & & & \\   
    \multicolumn{1}{c}{\resizebox{!}{0.4in}{\includegraphics{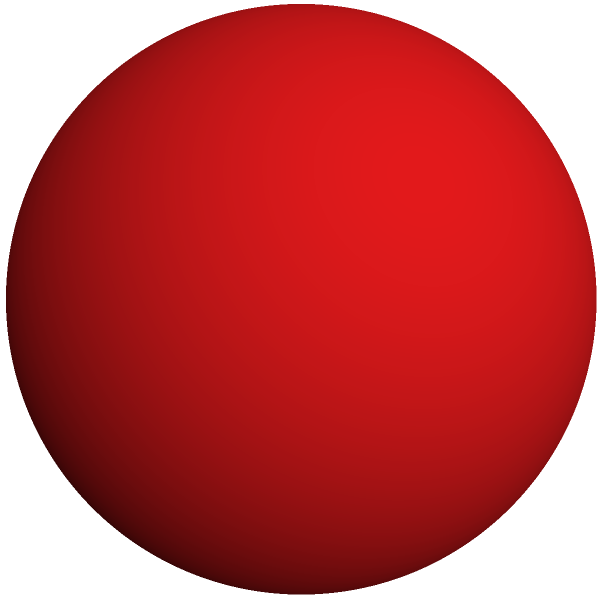}}} &
    \multicolumn{1}{c}{\raisebox{1.2\height}{\includegraphics[scale=0.03]{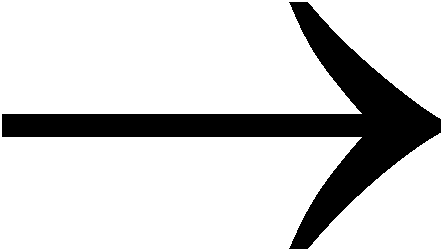}}}&
    \multicolumn{1}{c}{\resizebox{!}{0.4in}{\includegraphics{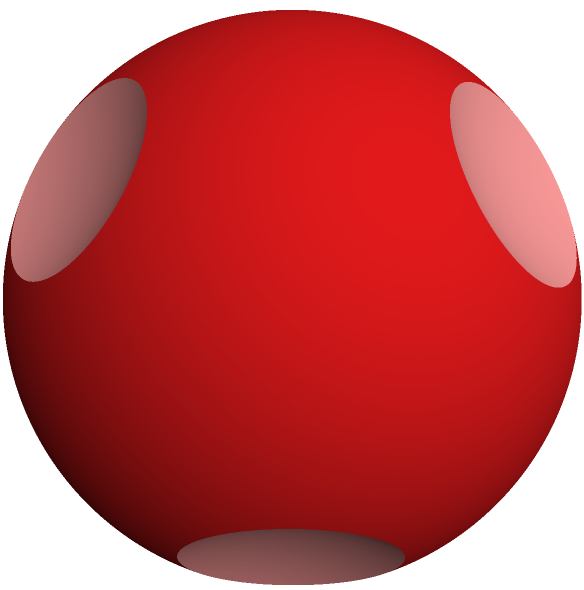}}} &
    \multicolumn{1}{c}{\raisebox{1.2\height}{\includegraphics[scale=0.03]{rightArrowTight}}}&
    \multicolumn{1}{c}{\resizebox{!}{0.4in}{\includegraphics{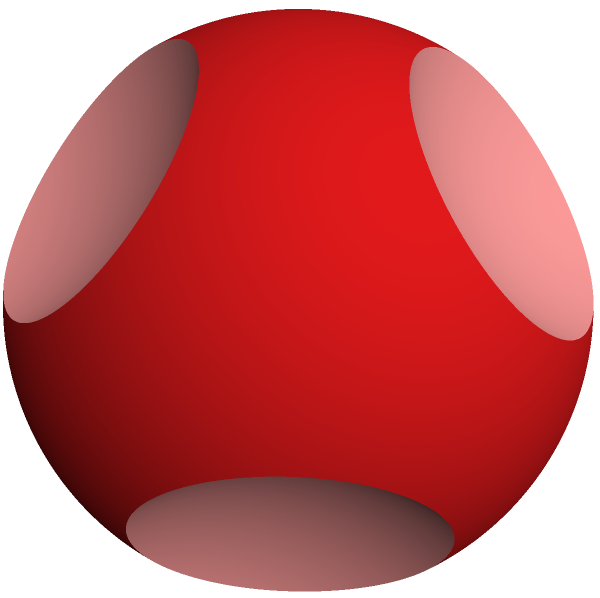}}} &
   \multicolumn{1}{c}{ \raisebox{1.2\height}{\includegraphics[scale=0.03]{rightArrowTight}}}&
   \multicolumn{1}{c}{ \resizebox{!}{0.4in}{\includegraphics{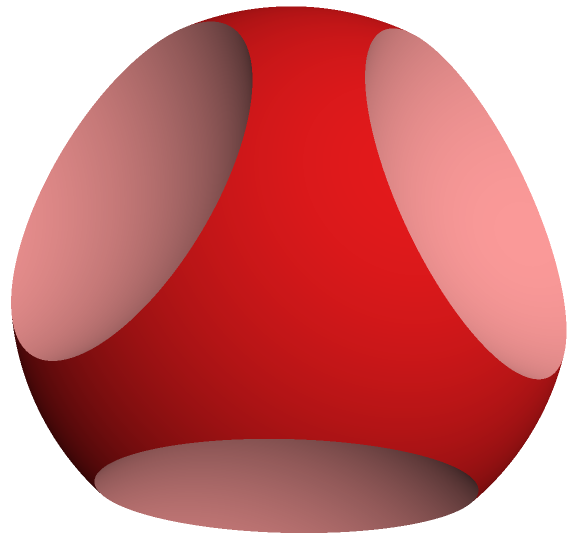}}} &
   \multicolumn{1}{c}{ \raisebox{1.2\height}{\includegraphics[scale=0.03]{rightArrowTight}}}&
   \multicolumn{1}{c}{ \resizebox{!}{0.4in}{\includegraphics{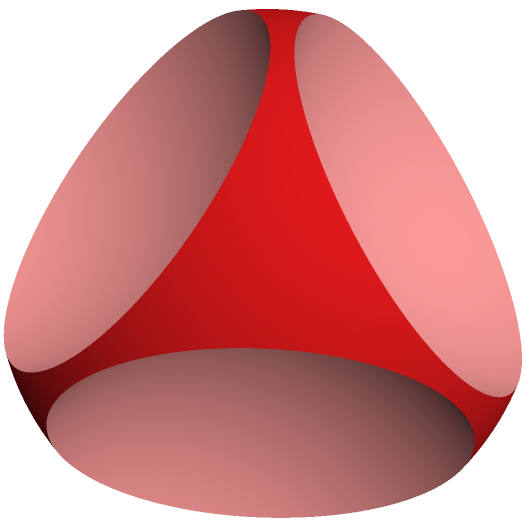}}} \\
    0.00 & & \multicolumn{1}{c}{0.25} & & \multicolumn{1}{c}{0.50} & & \multicolumn{1}{c}{0.75} & & \multicolumn{1}{c}{1.00} &  \\
\multicolumn{10}{c}{Dimpling Amount} \\
    \multicolumn{1}{l}{\textbf{(b)}} & & & & & & & &  \\
    \multicolumn{5}{c}{ \resizebox{!}{0.95in}{\includegraphics{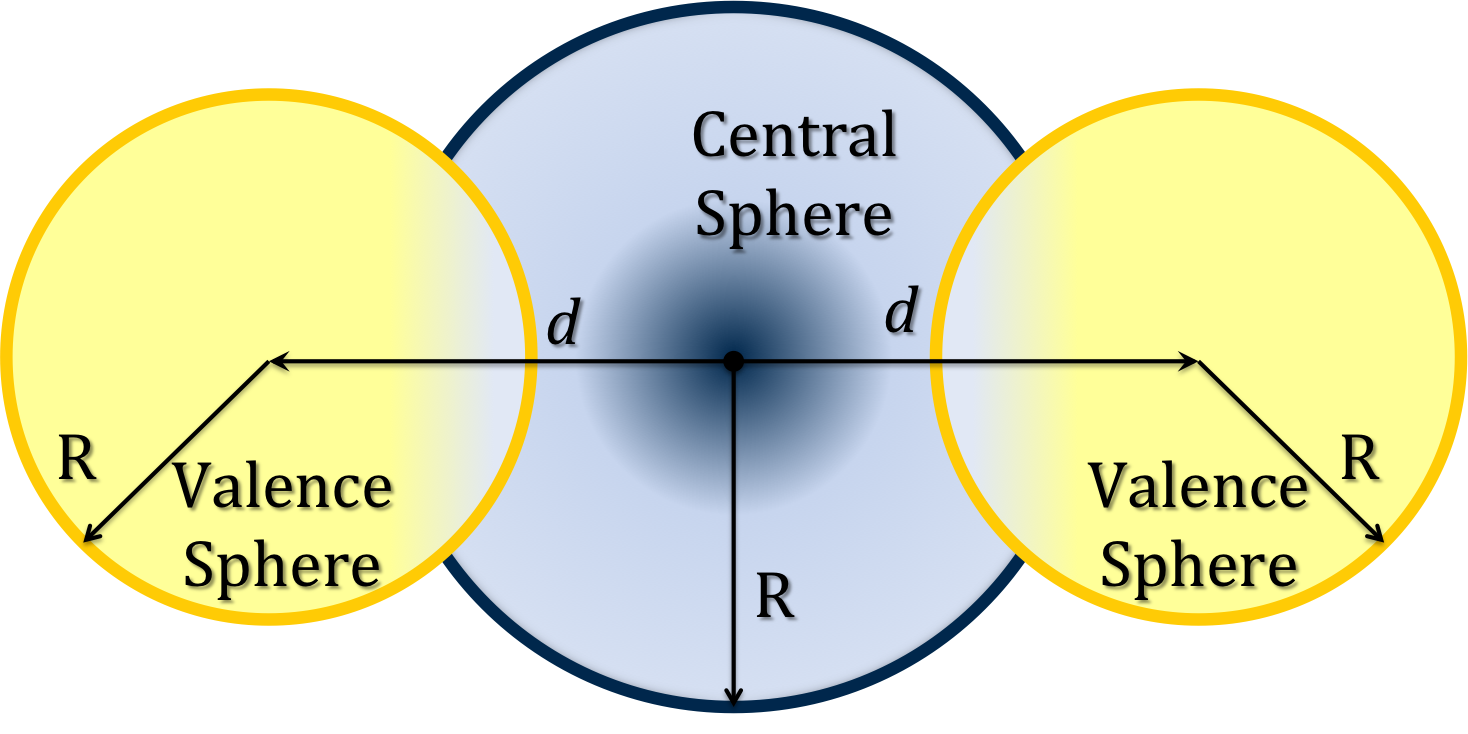}}} &
    \multicolumn{2}{c}{\raisebox{0.7\height}{\includegraphics[scale=0.09]{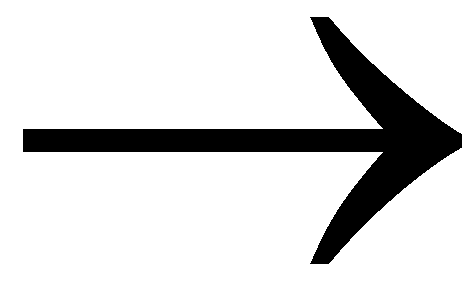}}} &
     \multicolumn{2}{c}{\resizebox{!}{0.9in}{\includegraphics{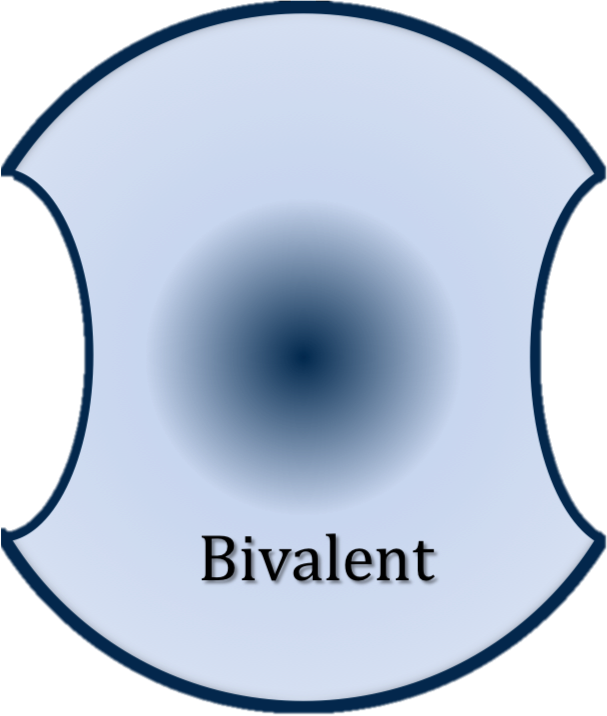}}}  \\
    \multicolumn{1}{l}{\textbf{(c)}} & & & & & & & & \\
   \multicolumn{1}{r}{ \resizebox{!}{0.58in}{\includegraphics{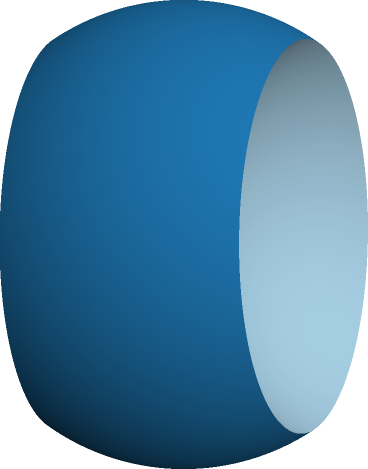}}} &
    \multicolumn{2}{c}{\resizebox{!}{0.58in}{\includegraphics{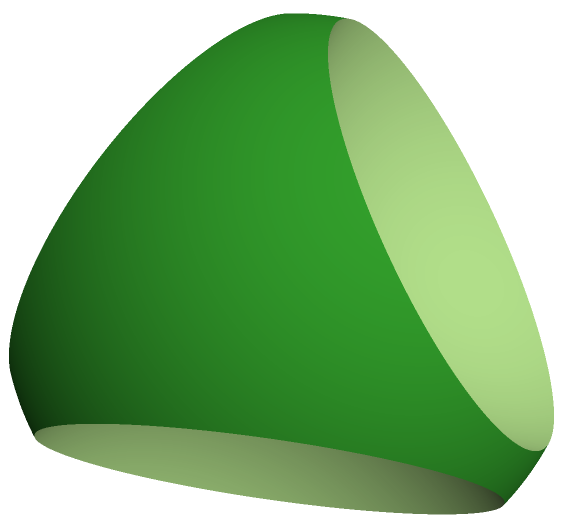}}} &
    \multicolumn{2}{c}{\resizebox{!}{0.56in}{\includegraphics{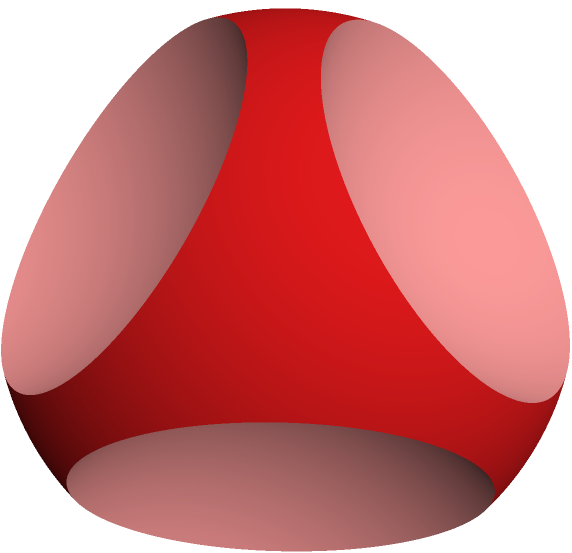}}}  &
    \multicolumn{2}{c}{\resizebox{!}{0.58in}{\includegraphics{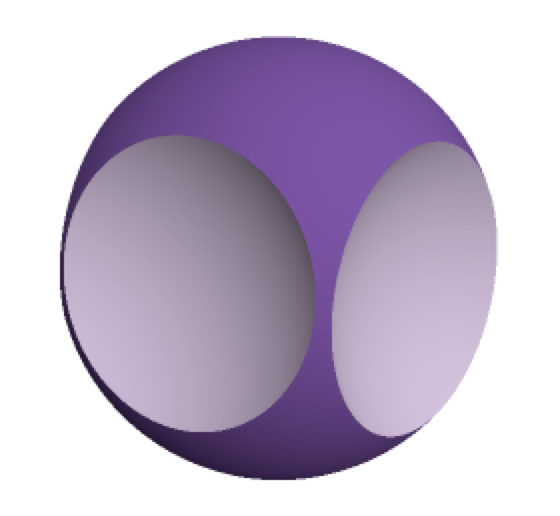}}} &
   \multicolumn{2}{c}{\resizebox{!}{0.54in}{\includegraphics{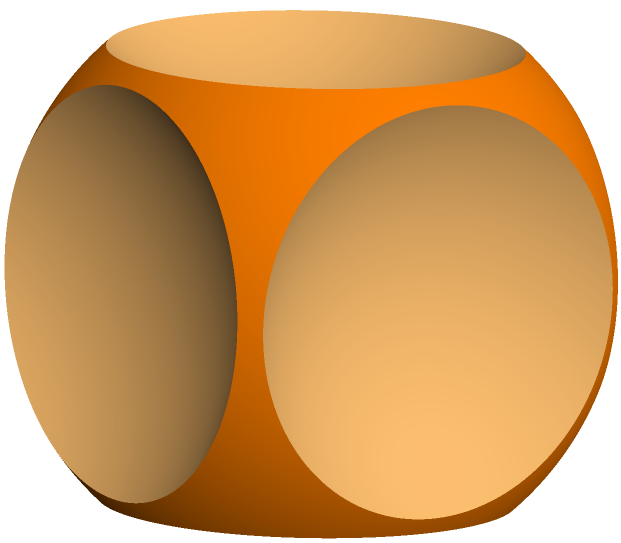}}} & \\  
    P2 & \multicolumn{2}{c}{P3} & \multicolumn{2}{c}{P4t} &\multicolumn{2}{c}{P4s} &
   \multicolumn{2}{c}{P6} &  \\
  \end{tabular}
  \caption{(a) Change in shape of a tetrahedral tetravalent particle from convex to concave as dimpling amount is increased from $0$ to $1$. (b) Parameterization of dimpled concave spheroidal particles. (c) Bivalent, trivalent, tetrahedral tetravalent, square tetravalent and hexavalent model spheroidal particles.}
  \label{fig:particles}
\end{figure}

Anisotropic colloids that exhibit emergent directional entropic forces
(DEFs)~\cite{entint} when crowded in a thermal system are entropically patchy
particles.\cite{epp} Entropic patches are shape features that promote
local packing into target motifs and drive systems to order globally at
high density.\cite{epp} Previously, we found that global ordering across a
variety of systems was consistent with entropic patch strengths on the order of
a few $k_\mathrm{B}T$.\cite{entint,epp} However, if we desire particles with
entropic patches of this strength, we do not currently have a complete set of
rules for understanding how the particle geometry needs to be controlled to get
them.\cite{digitalalchemy} Due to the emergent nature of the underlying
mechanism, establishing quantitative control is
difficult.\cite{entint,epp,digitalalchemy} Because of this difficulty, the rules
we previously proposed for controlling DEFs were heuristic.\cite{epp} Here, we
attempt to make the rules for controlling emergent DEFs more quantitative for a
specific case of families (see Fig.\ \ref{fig:particles}) of anisotropic
lock-and-key
colloids~\cite{Sacanna2010}.

Lock-and-key colloids are a convenient model system for investigating
how to use geometry to control DEFs. More importantly, a number of theoretical
\cite{Odriozola2008a,Marechal2010,Marechal2010a,Odriozola2013,Ashton2013} and
experimental investigations
\cite{Sacanna2010,Wang2014a,holeshell,stickydimple,stefanosynth,
dimpledisorder,Desert2013,sacannabuckle,colon2014} have shown that lock-and-key
colloids can form interesting structures, and recently synthesized families of
multiply dimpled particles \cite{Desert2013,Wang2014a} have not been previously
simulated. Here, we use computer simulation and analytic calculations to study both packing (infinite
pressure) and assembly (finite pressure) behaviors of families of
multiply-dimpled particles that differ by dimple number and symmetry as a
function of dimple size. We find that in both cases there are common,
approximate microscopic parameters, based on particle geometry, that correlate
with changes in bulk system behavior. However, the microscopic parameters are
different in the dense packing and assembly limits. In the dense packing limit,
a combination of analytic and numerical calculations show that (i) as dimple
size increases, there is a crossover from body-centered-cubic (BCC) or
-tetragonal (BCT) phases to phases that reflect the number of dimples and
particle symmetry, and surprisingly (ii) this point of crossover occurs
approximately independent of particle symmetry and number of dimples. Similarly,
in the assembly limit, we find that as dimple size increases in each family
there is (i) a crossover from the face-centered-cubic (FCC) structures that are
characteristic of hard sphere assembly~\cite{Hales2005} to BCC structures that
are characteristic of soft sphere assembly~\cite{Robbins1988}, followed by (ii)
a second crossover from BCC to behavior that is determined by the particle
symmetry and number of dimples.  Surprisingly (iii) this second point of
crossover also depends on a quantity that is approximately independent of
particle symmetry, though it is different from the control parameter in the
dense packing limit. We find that in the dense packing limit, the crossover in
behavior is determined by a quantity that depends simultaneously on the geometry
of all of the particle features, whereas in the assembly limit it is determined
by the volume of individual dimples. We interpret these results in light of
previous work~\cite{entint} arguing that shape entropy maximization causes
entropically driven colloidal systems to optimize local packing. Our results
provide detailed quantitative evidence on how shape features control the
macroscopic phase behavior of colloidal systems, while providing a concrete
example of a set of experimentally realizable systems exhibiting changes in
emergent macroscopic behaviors that are correlated with changes in a common
microscopic control parameter.

\section{\label{sec:model}Model and Methods}
\subsection{Model Systems}
Particles, shown schematically in Fig.\ \ref{fig:particles}(c) and denoted
by P$n_D$, are made of a central sphere P that is symmetrically dimpled by
subtracting $n_D$ valence spheres, all of the same radius as the central sphere
(hereafter taken to be one). The valence sphere positions are chosen to be
related by discrete symmetries, and are equidistant from the central sphere as
shown in Fig.\ \ref{fig:particles}(b). The use of equal radii
suggests a
connection between the dimple volume and free volume gained by the rest of the
system when particles bind entropically~\cite{epp,entint} due to shape
complementarity. We study five families of particles that satisfy two criteria:
they are multi-dimpled (so that we can differentiate between effects that are
driven by individual shape features and effects that are driven by overall
particle shape) and have discrete symmetries that are commensurate with some
lattice symmetry group. We denote the particles as: bivalent (P2), trivalent
(P3), tetrahedrally tetravalent (P4t), square tetravalent (P4s), and hexavalent
(P6). For each of these concave particles, there is a one-parameter family of
shapes determined by the distance $d$ between the central and valence spheres
which is maximum when the valence spheres are tangent to the central sphere and
minimum when the valence spheres are tangent to each other.

\begin{table}[ht] 
\caption{Particles from the Concave Spheres Family} 
\centering 
\begin{tabular}{c c c c c c} 
\multicolumn{5}{c}{ } \\
\hline\hline 
Particle & Description & Arrangement & Domain & Dimpling Amount\\ [0.5ex] 
\hline \hline 
P2 & Bivalent & Linear & $1 \leq d^2 \leq 4$ & $\frac{(4-d^2)}{3}$ \\ 
P3 & Trivalent & Triangular & $\frac{4}{3} \leq d^2 \leq 4$ & $\frac{3(4-d^2)}{8}$ \\ 
P4t & Tetravalent & Tetrahedral & $\frac{3}{2} \leq d^2 \leq 4$ &
$\frac{2(4-d^2)}{5}$\\ 
P4s & Tetravalent & Square & $2 \leq d^2 \leq 4$ & $\frac{(4-d^2)}{2}$\\ 
P6 & Hexavalent  & Cube & $2 \leq d^2 \leq 4$ & $\frac{(4-d^2)}{2}$\\ [0.5ex] 
\hline 
\end{tabular} 
\label{table:particles} 
\end{table}

Dimples are pairs of spherical caps bounded by the intersection of the
central and valence spheres, and have the following volume, surface area, and
circumference:
\begin{equation*}
  \begin{split}
    {v}_{d} &= \tfrac{\pi}{12} (d +4) {(2-d)}^2 \; , \\
    {s}_d &= \pi (2-d) \; , \\
    {c}_d &= \pi\sqrt{4-d^2} \; .
  \end{split}
\end{equation*}

\subsection{Overlap Algorithm for Dimpled Spheres}
Standard methods (see, e.g., \cite{gjk}) for determining the overlap between
anisotropic bodies rely on convexity. Previous theoretical work on
concave particles in
3D~\cite{Marechal2010,Marechal2010a,Cinacchi2010,Cinacchi2013} used
modifications of an overlap algorithm introduced by He and Siders~\cite{He1990}
that required the radius of the valence sphere to be smaller or equal to the
parent sphere and the particles were restricted to one valence sphere,
$(n_D=1)$. For the multi-dimpled particles we study here, we use a recently
developed overlap algorithm \cite{Beth2014}. The algorithm \cite{Beth2014}
is a recursive algorithm based on the
Cayley-Menger volume determinant and similar dihedral angle determinants.  This
overlap algorithm considers all topological types of overlap among the spheres
and calculates the mutual union and the mutual intersection among them.

\subsection{Dense Packing}
We vary the square
of the distance between the centers of the central and the valence spheres $d^2$
in increments of 0.01 for each of the particles, as shown in Table
\ref{table:particles}. The domain for $d$ varies between particles, so we
normalize by linearly mapping $d^2$ to the ``dimpling amount'' $f$, where
$f=0$ ($f=1$) when $d^2$ is maximum (minimum).

We construct dense packings of multi-dimpled particles numerically following a
standard protocol used in several other recent studies
\cite{Haji-Akbari2009,Chen2010,Damasceno2012,Damasceno2012a,Ortiz2014b,Chen2013}.
The densest packing structure is found by compressing systems with up to eight
particles placed in a box with periodic boundary
conditions~\cite{Chen2010,Chen2013}. Isobaric $(NPT)$ simulations, in which the
box is allowed to take the shape of an arbitrary parellelepiped, are then run
with a slowly increasing pressure. Each system is compressed $400$ times,
and the densest packing overall is recorded.

As has been done previously for particles of other shape,
\cite{Chen2010,Chen2013} where tractable, we use numerical putative densest
packings as \textit{ansatzes} for analytical packing constructions, which can be used to
compute exact dense packing curves.

We identify crystal structures by replacing each particle by a point at its
centroid as in several previous works
\cite{Zhang2005,Damasceno2012,Damasceno2012a,epp,Ortiz2014b}.

\subsection{Self Assembly}
We study the self assembly of multi-dimpled particles with MC
simulations using techniques from several other recent studies
\cite{Haji-Akbari2009,Chen2010,Damasceno2012,Damasceno2012a,Ortiz2014b,Chen2013,
epp,entint}. We use isochoric ($NVT$) MC simulations with periodic
boundary conditions to study systems of $N$ particles, with $N$ ranging from
$512$ to $2064$. These system sizes are typical of recent work examining entropy
driven self assembly of hard shapes (e.g.\
\cite{Haji-Akbari2009,Damasceno2012,Damasceno2012a,epp}) Each system is
initially equilibrated as a dense, metastable fluid for $10^5$ MC sweeps.
Isochoric ($NVT$) simulations are
then performed where the packing fraction is slowly increased until it reaches a
target value. The system is then allowed to equilibrate at this state point for
$\sim 5 \times 10^8$ MC sweeps. Equilibration times were chosen based on typical
equilibration times that have been observed in other studies of hard shapes~\cite{Haji-Akbari2009,Damasceno2012}, 
as their assembly originates 
in similar entropic forces. \cite{entint}
A few experiments were also performed to $\sim 10^{10}$ MC sweeps to 
verify that despite their concave shape, 
no nucleation or crystal growth is found to 
occur beyond typical equilibration times for hard convex particles.
Densities are varied between $0.25 \leq
\phi \leq \phi^*$, where $\phi^*$ is the densest known packing fraction for that
particle.  A full MC sweep for a system of $N$ particles consists of
$N+1$ trial moves including arbitrary translation, rotation, and box shearing
moves. Maximum step sizes are periodically adjusted to keep the acceptance
probabilities at 30$\%$. To check each self-assembled structure for
equilibration, we allow the system to evolve over time scales that are at least
an order of magnitude larger than the time scale for nucleation and growth, and
longer than system auto-correlation times.  All findings are verified by running
independent simulations with different initializations and in different box
sizes. The assemblies reported are those that form at the minimum density.

\section{\label{sec:results}Results and Discussion}
\subsection{\label{sec:DP}Densest Packings}
\begin{figure}
    \includegraphics[width=\columnwidth]{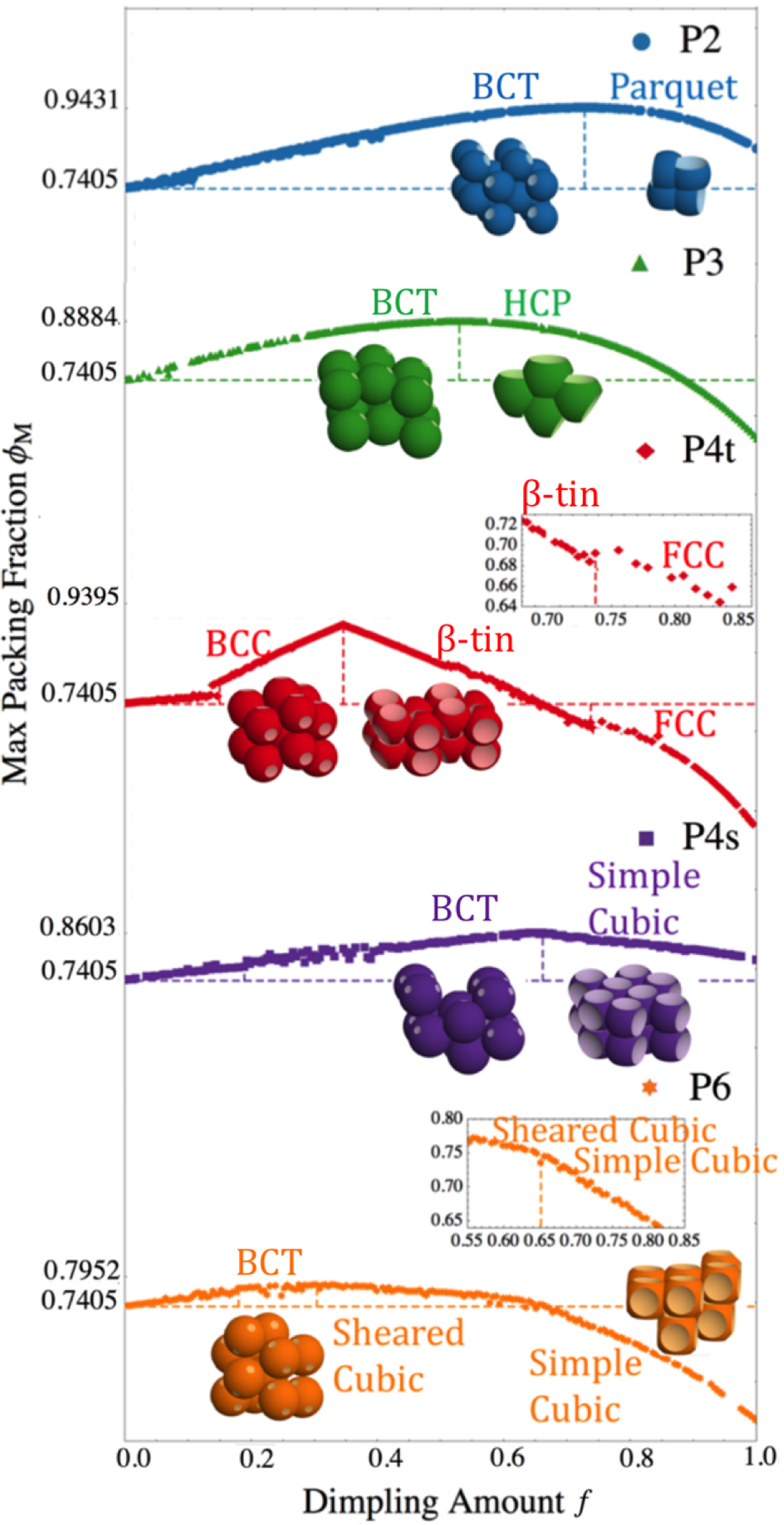}
  \caption{Numerical calculations of the density $\phi$ of the densest obtained
  packing for dimpled particles as a function of dimpling amount $f$. The curves
  for different particles are shifted along the y-axis for clarity. Packing
  fraction is maximum at a critical dimpling amount $f= f_{D\!P}^*$ when the
  dominating features switch from convex to concave regions of the particles.
  For small $f$, particles shear from a thermodynamic preference for FCC packing
  in the hard sphere limit to a BCT packing with the introduction of dimpling.
  This is not smoothly captured in numerical calculations, but we show that the
  transition is smooth through analytic packing calculations. (Right Insets) For
  P4t and P6 particles, a second transition at larger $f$ is seen from
  $\beta$-tin to FCC and sheared cubic to simple cubic respectively.}
  \label{fig:DPTrend}
\end{figure}

Following~\cite{Kallus2011,Damasceno2012,Chen2013}, we study continuous shape
deformations. We compute putative densest packings for each family of particles
at $100$ different dimpling amounts $f\in[0,1]$; see Fig.\ \ref{fig:DPTrend}.
To verify the putative densest packings predicted numerically
via MC simulations in this section, we perform analytic packing calculations for
our particles (see appendix \ref{analyticPackings}).  We find that our numerical
calculations match very well with the analytic findings, with an error less than
0.001\%.

\subsubsection{FCC--BCT Transition} 
In all cases, for small $f \approx 0$, particles pack most densely into FCC
lattices, like hard spheres~\cite{Hales2005}.  For slightly larger values $f
\leq f_{D\!P}^*$, all particle types pack like soft
spheres~\cite{Damasceno2012a} into a BCT structure. This transition from FCC to
BCT packing structures stems from a new contact between particles upon dimpling,
dimple-sphere contact, in addition to sphere-sphere contact. The dimple-sphere
contact reduces the contact distance between neighboring particles.
This results in a BCT structure with $\alpha = \beta \neq \gamma$, where
$\alpha$, $\beta$, and $\gamma$ are lattice distances.  The transition from FCC to
BCT structures occurs almost instantaneously with dimpling, and is captured in
our analytic densest packing calculations. We represent this with a
single continuous curve in our analytic calculations in Figs.\
\ref{fig:P2N_compare}-\ref{fig:P6N_compare}. However, in numerical putative
densest packing calculations, the particles pack in the FCC structure until a
large enough dimpling amount due to local frustrations.
At these values of $f$,
the dimples become increasingly filled by adjacent particles with decreasing
system volume, allowing denser packings without the need for global structural
rearrangement.

\subsubsection{Critical Dimpling Amount} 
We observe a maximum in the densest packing curve of each family at a different
critical dimpling amount $f^*_{D\!P}$.  In each family, below the critical
dimpling amount $ f < f^*_{D\!P}$ particle dimples make contact with the
convex part of adjacent particles. At $f > f_{D\!P}^*$, neighboring
particles interlock within each other resulting in a crossover to a structure
determined by the number and arrangement of dimples on individual particles.
Coincident with the change in the contact between particles is a
crossover in global structure from BCT structures to structures that depend on
the number and arrangement of the concave particle features. Despite an apparent
geometric similarity in the nature of the crossover, i.e.\ in each case the
crossover is from BCT to a symmetry dependent structure, the point at which the
crossover occurs, $f = f^*_{D\!P}$, varies by approximately a factor of 
two across particle types. This variation suggests that the dimpling amount $f$
is not a well-chosen parameter for understanding the origin of the structural
crossover. Instead, we find that the quantity
\begin{equation}
  {C}_{d} =  \frac{n_D {c}_{d} }{2\pi d} \, ,
\end{equation}
where $n_D$ is the number of dimples, $c_d$ is the dimple circumference, and $d$
is the dimple depth, is in the range $C_d^*\!=\!1.23\!\pm\!0.06$
at the crossover across particle types. We note that the dependence
of this quantity on $n_D$, as well as $c_d$ and $d$, suggests that the
crossover is driven by changes in the geometry of the entire particle, and
contrasts with what we find for self-assembly in section \ref{sec:SA} below.

\begin{figure}[h] \begin{tabular}{c c c c c c}
    \multicolumn{2}{c}{{\includegraphics[width=0.33\columnwidth]{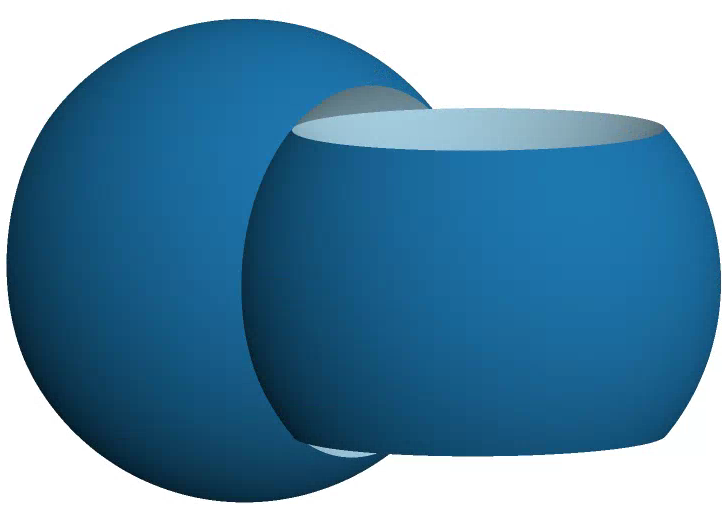}}} &
    \multicolumn{2}{c}{{\includegraphics[width=0.33\columnwidth]{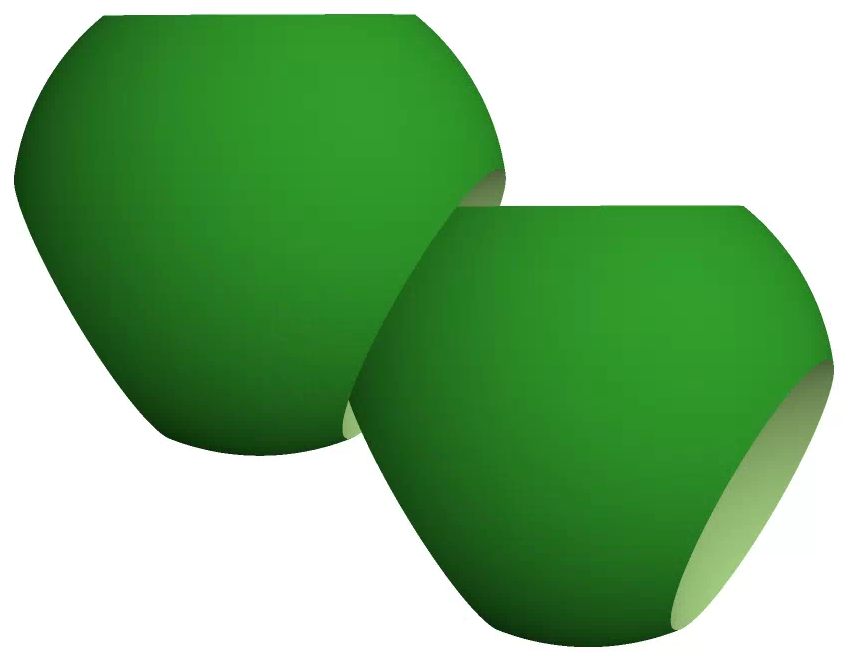}}} &
    \multicolumn{2}{c}{{\includegraphics[width=0.33\columnwidth]{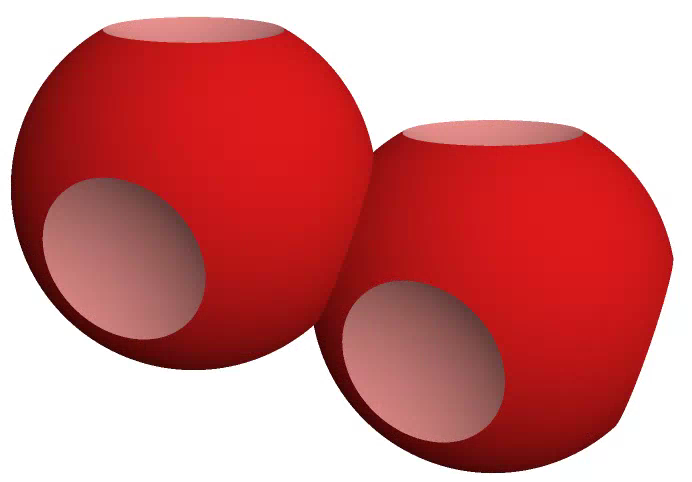}}} \\
    \multicolumn{2}{c}{(a) P2} & \multicolumn{2}{c}{(b) P3} &
    \multicolumn{2}{c}{(c) P4t} \\
    \multicolumn{3}{c}{{\includegraphics[width=0.33\columnwidth]{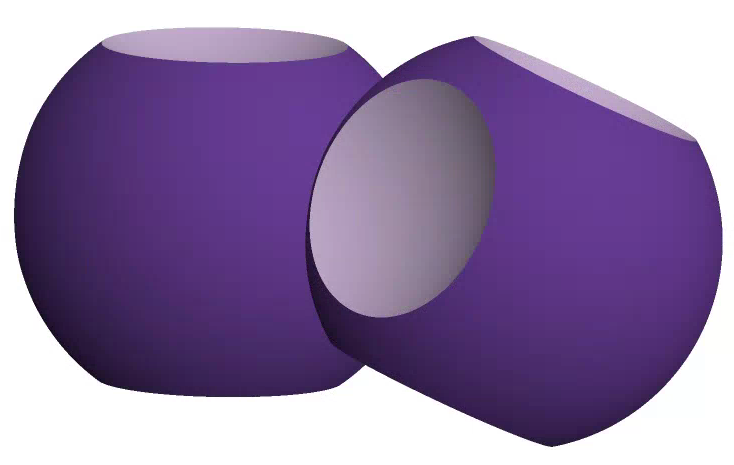}}} &
    \multicolumn{3}{c}{{\includegraphics[width=0.33\columnwidth]{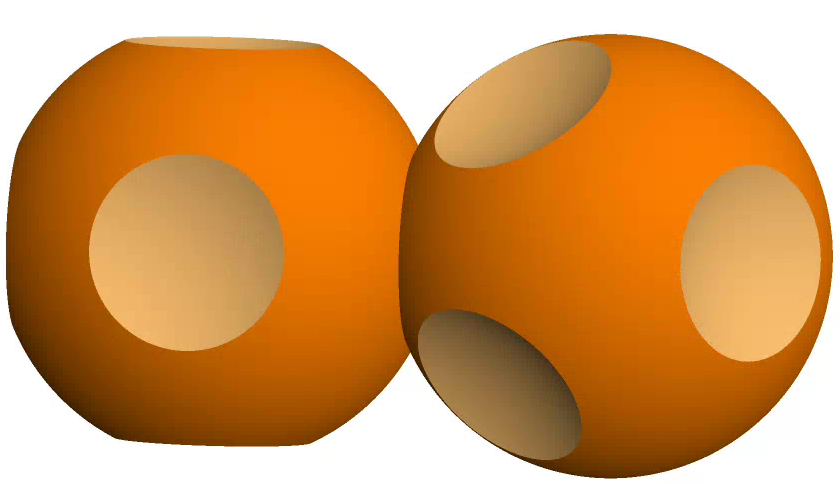}}} \\
    \multicolumn{3}{c}{(d) P4s} & \multicolumn{3}{c}{(e) P6} \\ \end{tabular}
    \caption{Configuration of neighboring particles at the critical dimpling
    amount for densest packing. We observe that the densest packing
    configuration is determined by a competition of the circumference and depth
    of the dimples.} \label{fig:criticalConfig} \end{figure}
 
We find that just above $f_{D\!P}^*$, P2 particles pack into a parquet structure
(BCC without shearing); P3 particles pack into an AB-stacked hexagonal packing
(hexagonal close packing--HCP); P4t particles pack into a $\beta$-tin lattice
(sheared diamond); P4s particles pack into a tetragonal lattice with rotational
parquet symmetry; and P6 particles pack into a simple cubic lattice. A further
increase in $f$ beyond $f_{D\!P}^*$ introduces larger voids between particles
and a reduction in packing fraction. As $f$ increases beyond $f_{D\!P}^*$,
dimple volume continues to increase, resulting in a smaller volume of the
particle in the unit cell structure. This change results in a reduction in the
packing density of these structures. At sufficiently large values of $f$, P3,
P4t, and P6 particles pack less densely than spheres.

For P2, P3, and P6 particles the packing fraction changes smoothly as
a function of $f$ about its maximum at $f_{D\!P}^*$; whereas for P4t and P4s
particles there is a cusp. Both types of behavior have been observed 
previously in dense packings of continuously varying shapes
\cite{Kallus2011,Damasceno2012,Chen2013}. For P2, P3, and P6 particles, the
smooth behavior occurs because structures on either side of $f_{D\!P}^*$ are
related by a continuous shear. For P4t particles the cusp is the result of a
crossover from BCC to $\beta$-tin, and for P4s it is the result of a change from BCC
to simple cubic with a rotational parquet symmetry in the simple cubic lattice,
which cannot be obtained by a simple shear from the BCC structure.

For P4t particles at high dimpling amounts, $f\!=\!0.7374$, we find another
transition from $\beta$-tin to FCC, shown in the first right inset of Fig.\
\ref{fig:DPTrend}. At these large dimpling amounts, the densest packing
structure arises from a competition between the parallel and anti-parallel
alignment of the dimples. With increased dimple size, neighboring
particles find more room to rotate while they are interlocked. This results in
a denser packing when the dimples align parallel to each other instead,
resulting in a transition from $\beta$-tin, where dimples exhibit anti-parallel
alignment, to the FCC structure. 

Similar to P4t particles, we find another transition for P6 particles at
$f\!=\!0.6510$, shown in the second right inset of Fig.\ \ref{fig:DPTrend}. The
particles transition from a sheared cubic arrangement to a simple cubic
arrangement. This transition is observed in hexavalent particles because of
their ability to shear along the {110} lattice vector direction, while
neighboring particles remain interlocked in the same configuration.

\begin{figure}[h]
  \includegraphics[width=\columnwidth]{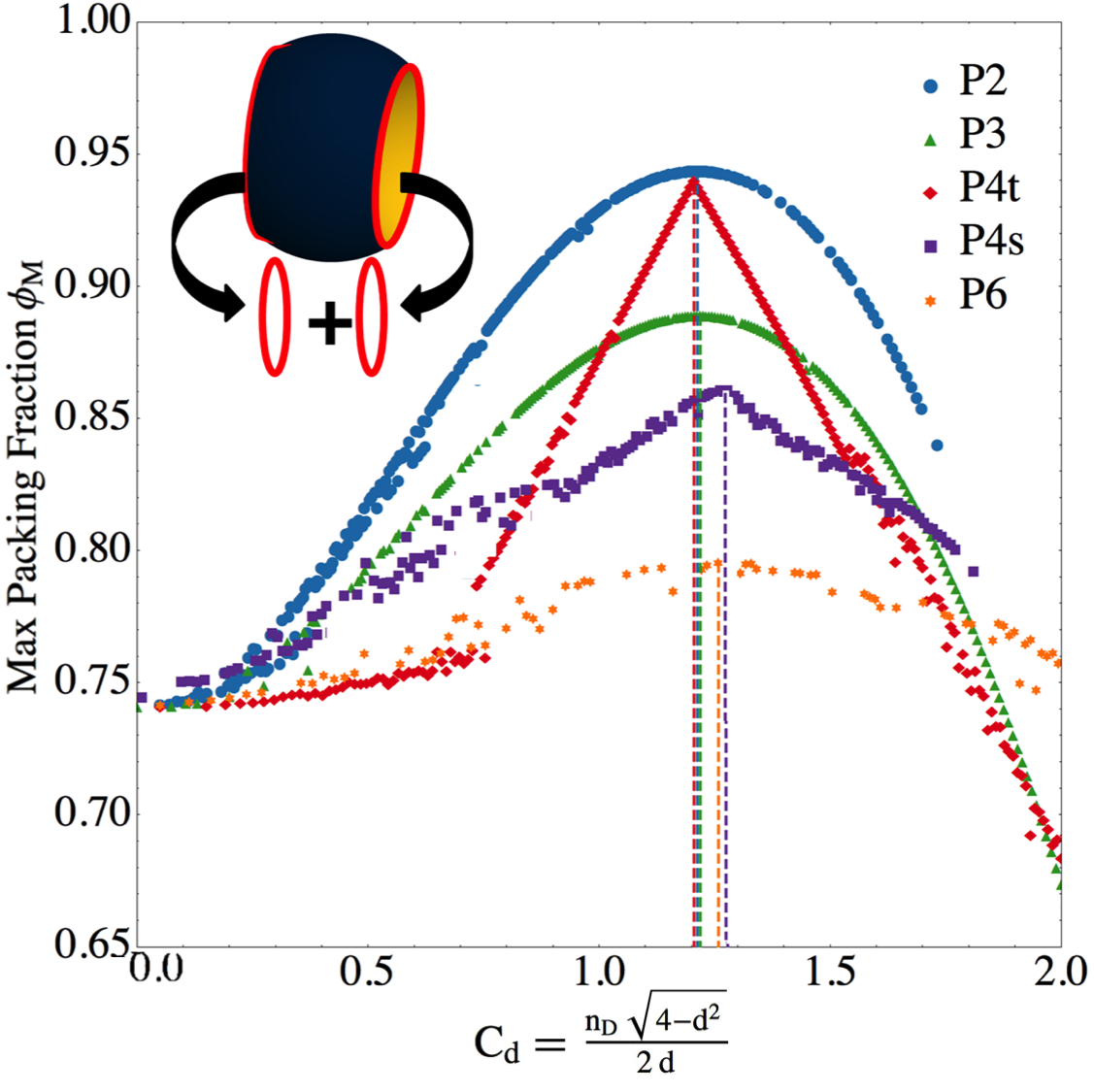}
  \caption{Packing fraction $\phi$ vs the total dimple circumference-depth ratio
  $C_d$. The packing fraction reaches a critical maximum at
  $C_d^*\!=\!1.234\!\pm\!0.060$.}
  \label{fig:scaling_DP}
\end{figure}

\subsection{\label{sec:SA}Self-Assembly}
We also performed MC simulations at finite pressures to study assembly behavior
away from the dense packing limit.

\subsubsection{Symmetry Independent Behavior near $f=0$: Hard to Soft Transition}
\label{sec:SAHardSoftTrans}

\begin{figure}
  \includegraphics[width=\columnwidth]{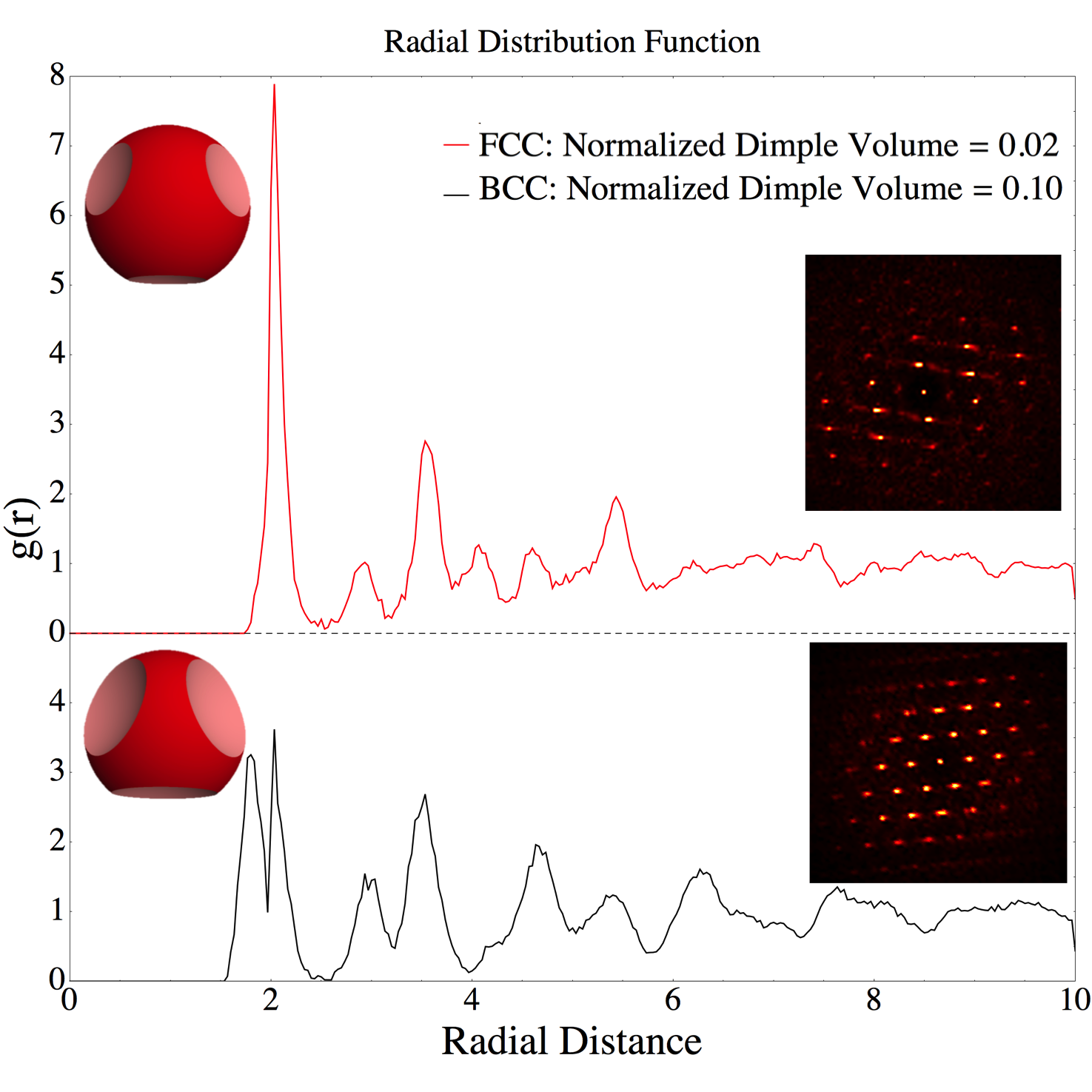}
  \caption{The radial distribution function of tetrahedrally tetravalent
  particles (P4t) at infinitesimally small dimple volume (above) and small
  dimple volume (below). We report a transition from a face-centered cubic to a
body-centered cubic crystal structure as volume of the dimple increases.}
  \label{fig:FCC_BCC_trans}
\end{figure}
In all cases, at $f=0$ we observe the assembly of FCC lattices as we would
expect for hard spheres. In all cases, as dimpling increases from $f=0$ we
observe a crossover from FCC to sheared BCT structures.  The onset of the BCT
structure occurs almost instantaneously as a result of continuous shearing from
FCC into BCC structures.\cite{Dunlap2012} The FCC-BCT transition is followed by
a transition from BCT to BCC at dimple volume $v_d \sim 0.03$, which can, again,
occur through simple shearing of the lattice.\cite{Dunlap2012} The self-assembly of BCC
structures from nearly spherical hard particles has been observed previously
(see, e.g., \cite{Damasceno2012a,epp}).  Here, by means of the radial distribution
function, $g(r)$, (Fig.\ \ref{fig:FCC_BCC_trans} shows an example for P4t
particles) we see that in the BCC phase, neighboring particles penetrate into
the dimples, but potential of mean force and torque (PMFT) calculations (Fig.\
\ref{fig:pmft_panel}a,c,e and Fig.\ \ref{fig:pmft_panel2}a,c) show that,
overall, the distribution of nearest neighbors remains relatively isotropic
despite the presence of dimples. Other work \cite{Robbins1988,Monovoukas1989}
has shown that soft spheres assemble BCC and other non-close packed lattices. In
light of this prior work, our radial distribution function and PMFT results
suggest that the BCC phases we observe here for nearly spherical particles, as
well as those observed elsewhere (e.g.\ \cite{Damasceno2012a,epp}), arise
because nearly spherical hard particles behave effectively as soft spheres,
regardless of their symmetry.

\subsubsection{Symmetry Dependent Behavior at Large Dimpling}
\label{sec:ConcaveSA}

\begin{figure}
    \includegraphics[width=\columnwidth]{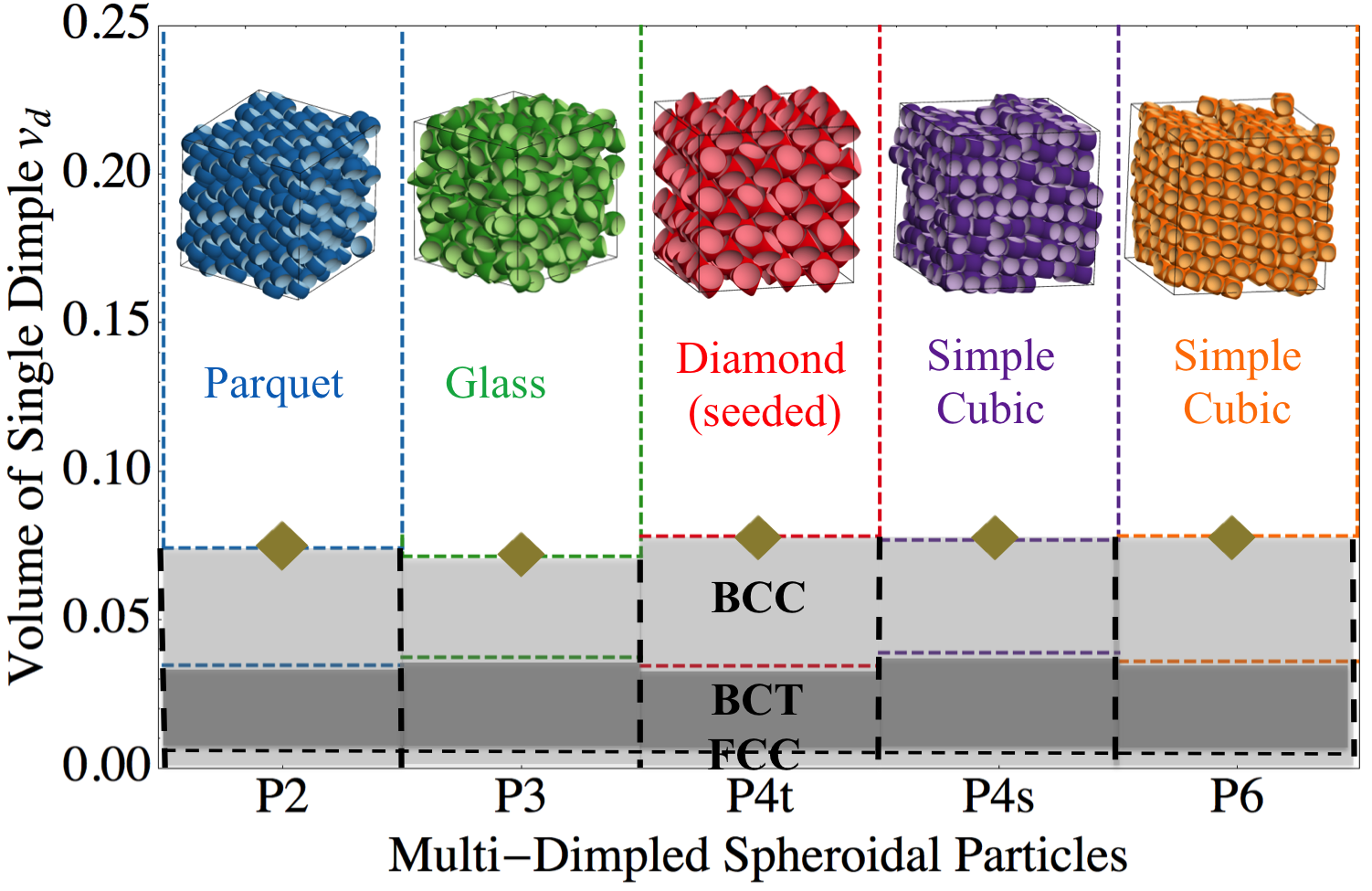}
  \caption{Self-assembled crystals for different multi-dimpled spheroidal particles vs volume of individual dimples in particles. We find that critical volume of an individual dimple determines the spherical ($v_d \le 0.07$) or non-spherical self-assembly of these particles. We also observe that hard spheres ($v_d = 0$) assemble into the FCC lattice. With the introduction of an infinitesimally small dimple, the sheared body-centered tetragonal structure is found. Above $v_d \ge 0.03$, we find that all particles assemble into the BCC lattice.Error bars are within the size of the gold markers.}
  \label{fig:SA_scaled}
\end{figure}

Above a critical value $f \geq f_{S\!A}^*$ for each family, the phase behavior
is dependent on the arrangement of dimples, as shown in Fig.\
\ref{fig:SA_scaled} (for futher details, see appendix Figs.\
\ref{fig:P2N_TransD}-\ref{fig:P6N_TransD}). P2 particles continue
to self-assemble a BCC lattice, but the lattice takes on a parquet form in which
particle orientations are ordered (see section \ref{Appendix:PMFT}).  For P3 and
P4t we do not observe spontaneous assembly of an ordered structure from a
homogeneous fluid at any density on the time scale of our simulations.  For P3,
we find the expected triangular lattice is stable against melting to a packing
fraction of $0.35$ at dimpling amounts greater than $0.45$.  For P4t, if seeded,
particles assemble a diamond lattice as found previously for tetrahedrally
patterned enthalpically patchy particles~\cite{Zhang2005}, but unlike truncated
tetrahedra~\cite{Damasceno2012} or tetrahedrally facetted spheres~\cite{epp},
which assemble without seeding.  The diamond lattice does not melt down to a
packing fraction of $0.26$ at dimpling amounts greater than $0.60$. P4s
particles self-assemble a cubic lattice with rotational parquet symmetry. P6
particles self-assemble a simple cubic lattice. In all cases we observe that the
crossover from BCC coincides with an emergent valence, or angular specificity,
in the effective interactions between neighboring particles, on the order of a
few $k_\text{B}T$ (see section \ref{Appendix:PMFT}), as has been observed for
convex particles~\cite{entint}.

\subsubsection{Potential of Mean Force and Torque}
\label{Appendix:PMFT}
\begin{figure*}
\includegraphics[width=6.0in]{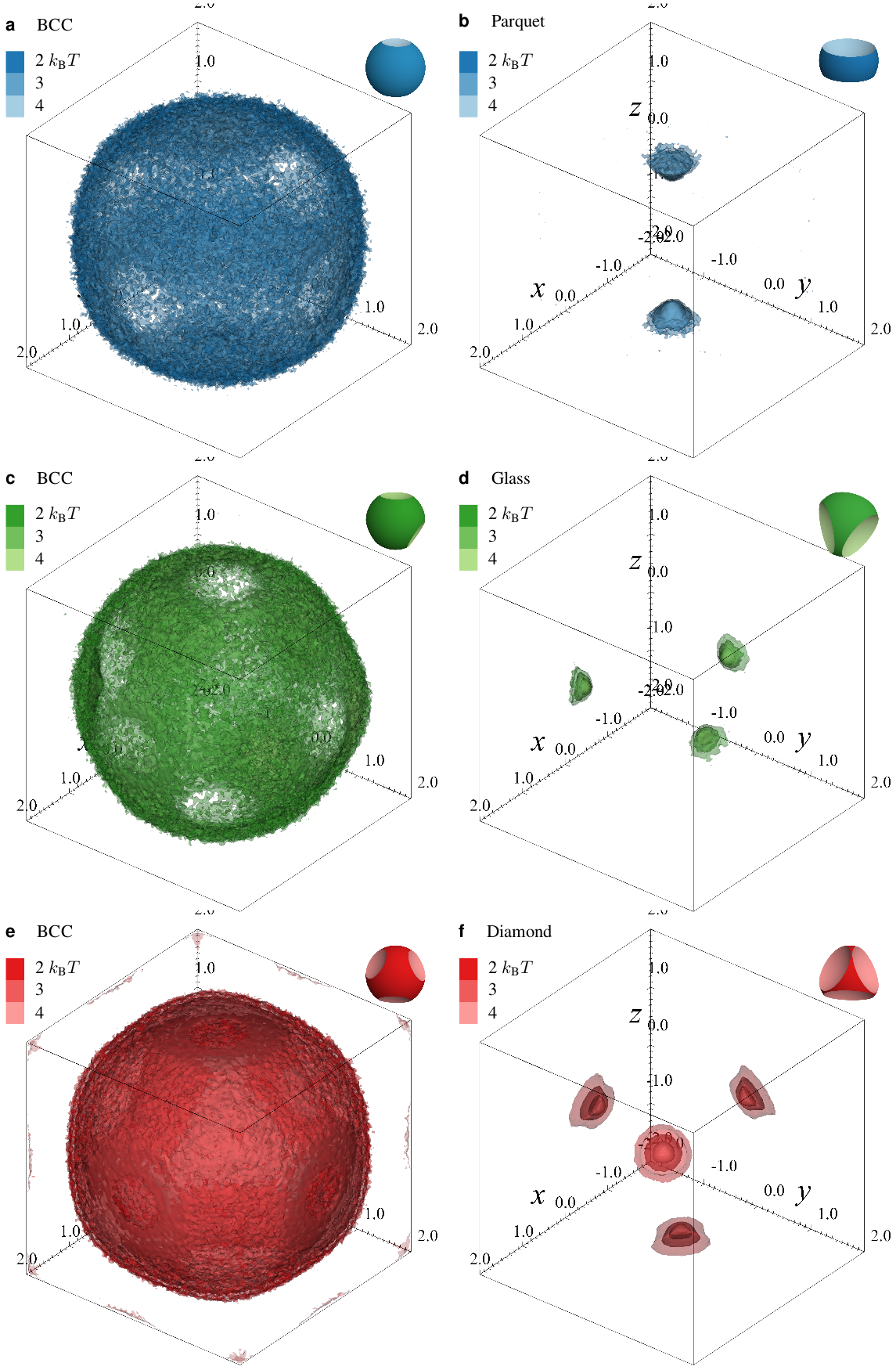}
\caption{Potential of mean force and torque (PMFT) calculations~\cite{entint}
show that below the critical dimpling amount $f_{S\!A}^*$ (left) an isotropic
potential exists. At higher dimpling amount (right), an attractive free energy
well is presented in the volume of the dimple, giving rise to different crystal
structures.  (a, b) Bivalent particle showing (a) BCC and (b) parquet
potentials. (c, d) Trivalent particle showing (c) BCC and (d) triangular sheet
potentials. (e, f) Tetrahedrally tetravalent particle showing (e) BCC and (f)
tetragonal diamond potentials.}
\label{fig:pmft_panel}
\end{figure*}

\begin{figure*}
\includegraphics[width=6.0in]{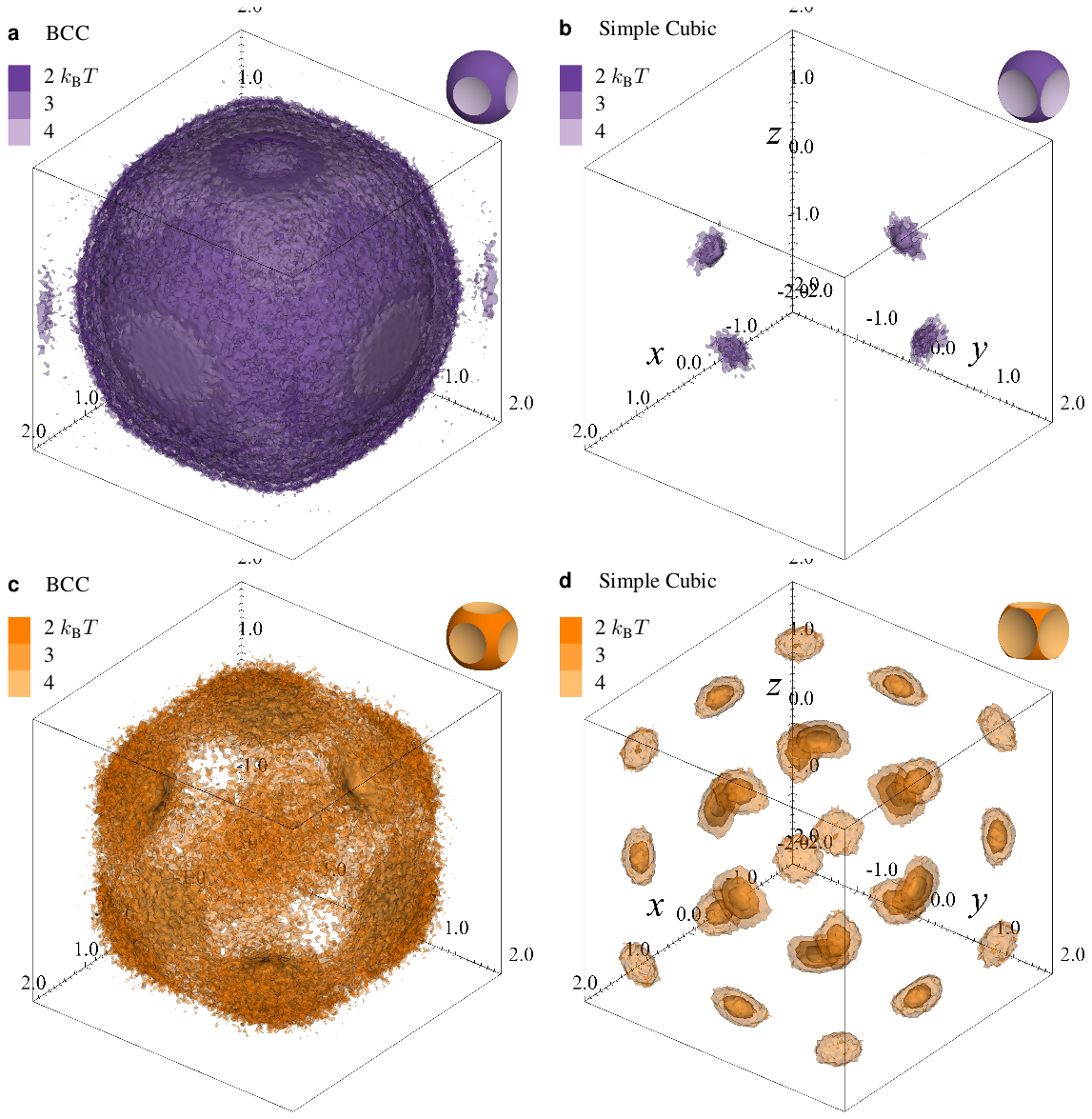}
\caption{Potential of mean force and torque (PMFT) calculations~\cite{entint}
show that below the critical dimpling amount $f_{S\!A}^*$ (left) an isotropic
potential exists. (a, b) Square Tetravalent particle showing (a) BCC and (b)
simple cubic potentials. (c, d) Hexavalent particle showing (c) BCC and (d)
simple cubic potentials. As the lattice vector of the cubic lattice formed is
small in this case, we see second and third neighbor peaks in the cubic
potential.}
\label{fig:pmft_panel2}
\end{figure*}

The PMFT is a measure of the effective directional entropic forces (DEFs) that
emerge for a pair of crowded particles \cite{entint}. The PMFT (Figs.\
\ref{fig:pmft_panel}, \ref{fig:pmft_panel2}) shows how the local particle
environment of multi-dimpled lock-and-key colloids differs in different bulk
structures. We computed the PMFT at the lowest density where the crystal
structure is formed, around a constant volume with a single particle placed at
the center using methods introduced in our previous works \cite{epp,entint}. In
the case of P3 and P4t particles, where no crystal structures are formed, we use
the lowest stable density of the crystal structure obtained from melting studies
of the densest packing structure. Figs.\ \ref{fig:pmft_panel},
\ref{fig:pmft_panel2} show isosurfaces of the PMFT in three dimensions plotted
with VisIT~\cite{HPV:VisIt}. In both columns, the PMFT is computed on the same
constant volume. Below the critical dimpling amount $f_{S\!A}^*$, we observe a
nearly isotropic effective entropic potential around the particle. At higher
dimpling amounts, the dimples behave as an entropic patch \cite{epp} that
results in a preferred directionality between pairs of crowded particles,
similarly to what was observed previously for facetted spheres
\cite{epp,entint}.

For large dimples, the PMFT (Figs.\ \ref{fig:pmft_panel}b,d,f and
\ref{fig:pmft_panel}b,d) show that a particle A adjacent to a dimpled particle B
will coordinate at a dimple of particle B. This is similar to what has been
observed previously for facetted spheres and polyhedra \cite{epp,entint}.
However, in contrast to what has been observed previously for other shapes,
depending on the structure, particle A will either preferentially align its
convex region with the dimple of particle B, or will align its dimple with the
dimple of particle B. This is a new feature that arises from the particle
concavity. However, even in the case where both patches on neighboring particles
align, the behavior differs from what has been observed previously. In the
diamond crystal structure obtained from tetrahedrally tetravalent (P4t)
particles, the dimples align facing each other, similarly to what has been
observed previously \cite{epp,entint}. However, because of the concave nature of
the entropic patches, there is considerable empty volume between the particles,
resulting in stability at lower packing fraction of $\phi=0.32$ than is evident
from the structure, and what has been observed for convex particles
\cite{epp,Damasceno2012}.

\subsubsection{Crossover in Self-Assembly from Symmetry Independent to Symmetry
Dependent Behavior}
As in dense packing, with increasing dimple size, in each family we observe a
change in system behavior. As we noted above, in all cases below some
$f_{S\!A}^*$, we observed the self-assembly of BCC phases. Above $f_{S\!A}^*$,
we observed phase behavior that depended on particle symmetry. The fact that in
all cases we observed a loss of rotational entropy (captured in the PMFT results
in Figs.\ \ref{fig:pmft_panel}, \ref{fig:pmft_panel2}) driven by changing the
particle geometry, suggests that there may be a common underlying geometric
origin of this change in behavior. Depending on particle symmetry $f_{S\!A}^*$
varied between $0.42$ and $0.70$ ($66.6\%$), which suggests that, as in dense
packing, $f$ is not a good parameter for characterizing the origin of the
crossover in bulk behavior. Instead, we found that the volume of an individual
dimple on a multi-dimpled particle showed relatively small variation ($0.0714
\le v_d \le 0.0783$, or $\Delta v_d/v_d \approx 10$\%) at the crossover in bulk behavior across different
particle symmetries.

We found that the change in bulk behavior coincides with the binding of
particles at dimples, and that this coincides with an approximately constant
volume of individual dimples. To understand why this might be the case, we note
that Ref.\ \cite{entint} argued that the phase behavior of several systems of
convex particles was driven by local packing considerations, which in example
systems could be reduced to the consideration of simple, pairwise packing.  Our
results suggest the argument advanced in Ref.\ \cite{entint}, i.e.\ that
pairwise packing considerations are sufficient to understand the phase behavior
of certain convex particles, extends to multi-dimpled sheres. We hypothesize why
this is the case by considering what happens when a particle A binds
entropically at a dimple of particle B. When particle A binds at the dimple of
particle B, particle B loses rotational entropy on the order of $k_\mathrm{B}
\ln \Omega$ (where $\Omega$ is the number of distinct particle
orientations\footnote{In technical terms, $\Omega$ is given by the volume of the
relevant rotation group.}). This entropy loss is compensated by a gain in
entropy for the rest of the system. It was argued in Ref.\ \cite{entint} that
this should be determined by the stress tensor, which we can estimate as
typically on the order of $k_\text{B}T/v_p$ where $v_p$ is the particle volume.
This suggests that the additional entropy gained by the system is on the order
of $k_\text{B} v_d/v_p$, where, as particle pairs bind, we assume that the rest
of the system gains free volume proportional to the dimple volume. There are
key differences between systems: the pressure at the onset of ordering, the
number of non-degenerate rotations (which is determined by the discrete symmetry
of the particle), and the orientation of particles coordinated at dimples, it
would be surprising to find perfect collapse for any geometric masure of
ordering. Nevertheless, we find the variation in critical dimple volume is
small.

We expect the result of dimple volume as the controlling feature for anisotropic
assembly is specific to multi-dimpled lock-and-key colloids, and we do not
expect this to be a phenomenon that generalizes to all
shapes, e.g.\ to colloidal ellipsoids. Indeed prior work on facetted spheres
suggested there that facet edges were important, but that was based on fewer
examples than the present work \cite{epp}. However, what we have found is that
the suggestion that the phase behavior of anisotropic colloids is controlled by
pairwise effective interactions, as argued in Ref.\ \cite{entint}, extends
beyond the convex particles studied in \cite{entint}, and suggests that the
argument advanced in that work may be universal, even if the quantitative
details of how it plays out for particles of different types is not.

\section{\label{sec:Conclusion}Conclusion}
We found a structural crossover in packing that corresponds to a quantity that
depends on \emph{all} features of a particle shape, and one for the local
ordering in assembly that depends on \emph{individual} particle features. This
finding is interesting in light of several studies showing discrepancies between
the packing and assembly behavior of hard particles (e.g.\
\cite{Haji-Akbari2009,escobedo,Damasceno2012,Damasceno2012a}).  Recent
work~\cite{entint} has argued that the discrepancies between packing and
assembly arise because assembly is driven by directional entropic forces that
arise from balancing shape entropy, which in turn drives \emph{local packing}
whereas packing at infinite pressure is a global phenomenon in which entropy is
irrelevant.  Our finding, that the local ordering at low pressures where we
observe assembly (to within 10\%) controlled the geometry of individual particle
features, whereas packing behavior is controlled (also to within 10\%) by the
geometry of all particle features, further supports the argument advanced in
\cite{entint}.

Interestingly, multi-dimpled particles self-assemble porous crystal structures
at packing fractions of $40 - 50\%$ (and under certain conditions can remain at
least metastable to much lower packing fractions), lower than that observed in
convex systems (typically $\geq 50\%$
\cite{DeGraaf2011,escobedo,Damasceno2012,Damasceno2012a,epp}). It has been
proposed recently that entropy can stabilize open lattices in systems with both
enthalpic and entropic contributions to the free energy~\cite{Mao2013}. Here, we
found that entropy alone can stabilize porous structures. Intuition might
suggest that the absence of strong enthalpic interactions in the present system
may also result in fewer kinetic traps, and in several cases we found robust
bulk 3D assembly. However, the kinetics of lock-and-key colloids are subtle: it
has been shown that in the presence of depletants single-dimpled particles first
make convex-to-convex contact before particles rearrange into convex-to-concave
contact \cite{colon2014}. The existence of this complex kinetic pathway may
explain why we did not observe the spontaneous global ordering of P3 or P4t
particles here, despite those particles having DEFs in the range that causes
other systems to self-assemble.  Despite these exceptions, porous structures of
the type we report here find applications in advanced catalysis and medical
diagnosis~\cite{Davis2002}. The application of recently developed techniques for
optimizing shapes for structures \cite{digitalalchemy} to multi-dimpled
particles might be useful for producing shapes that are less likely to get
kinetically trapped.

Finally, we have described a set of systems that exhibit complex emergent,
entropy-driven phase behavior that can be traced to a microscopic quantities
that are approximately invariant under changes of particle symmetry.  Because
emergent behaviors are inherently difficult to ascribe to microscopic system
details, we believe this example in experimentally realizable systems of
colloids \cite{Desert2013,Wang2014a} will provide a useful experimental setting
for manipulating emergent behavior.

\begin{acknowledgments}
We thank J.\ Crocker, M.\ Engel, and O.\ Gang for helpful discussions, and A.\
Karas for help with PMFT visualization. The U.S.\ Army Research Office under
Grant Award No.\ W911NF-10-1-0518 and the DOD/ASD(R\&E) under Award No.\
N00244-09-1-0062 supported the assembly work. The Biomolecular Materials Program
of the Materials Engineering and Science Division of Basic Energy Sciences at
the U.S.\ Department of Energy under Grant No.\ DE-FG02-02ER46000 supported the
packing studies and PMFT calculations. ERC acknowledges NSF MSPRF grant
DMS-1204686.
\end{acknowledgments}

\appendix

\section{Geometric Characteristics of Multi-Dimpled Particles}
\label{Appendix:Geometries}

\begin{figure}
 \begin{tabular}{c c c}
    \includegraphics[width=0.33\columnwidth]{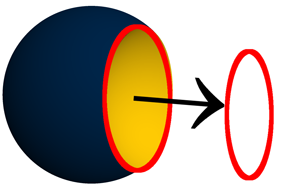} &
    \includegraphics[width=0.33\columnwidth]{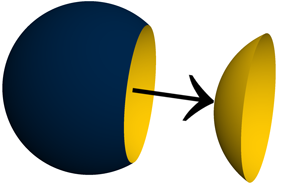} &
    \includegraphics[width=0.33\columnwidth]{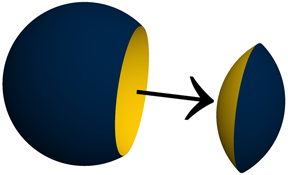} \\
    (a) Circumference &
    (b) Surface Area &
    (c) Volume \\
  \end{tabular}
  \caption{The different shape features of multi-dimpled spheres considered in
  this study.}
  \label{fig:shapeCharacTrend}
\end{figure}

We calculate geometric characteristics of the particles at various dimpling amounts, as shown in Figure \ref{fig:shapeCharacTrend}. 

\subsection{Dimple Volume}

The volume of a single dimple, ${v}_{d}$, is the union of the volume of the two
intersecting spherical caps that form the dimple. The volume of spherical cap of
radius $R$ and height $h$ is given by
\begin{equation*}
V(R, h) = \tfrac{1}{3} \pi h^2 (3R - h)\,.
\end{equation*}
For a single dimple formed at the intersection of central and valence spheres of
radii $R_+ , R_-$, its volume is given by
\begin{multline}
{v}_d = \tfrac{\pi}{12d} {(R_{+} + R_{-} -d)}^2\times\\
(d^2 + 2dR_{+} + 2dR_{-} + 6R_{+}R_{-} - 3R_{+}^2 - 3R_{-}^2)\end{multline}\,.
For $R_{+} = R_{-} = 1$,
\begin{equation}
{v}_{d} = \tfrac{\pi}{12} (d +4) {(2-d)}^2 \,.
\end{equation}
where $d$ is the distance between the centers of the central sphere and the
valence sphere.  The volume of the remaining particle ${v}_{s\!p}$ is given by
\begin{equation}
{v}_{s\!p} = {v}_{P\!S} - (n_D \times {v}_{d})\,.
\end{equation}
where ${v}_{P\!S}$ is the volume of the original central sphere, ${v}_{d}$ is
the volume of a single dimple and $n_D$ is the number of dimples in the
particle. This gives
\begin{equation}
{v}_{s\!p} = \tfrac{4}{3} \pi R_+^3 -  n_D \times \tfrac{\pi}{12}  (d+4) {(2-d)}^2 \,.
\end{equation}
We normalize the total dimple volume for a particle by 
\begin{equation}
{V}_{d} = \frac{n_D \times {v}_{d} }{v_{s\!p}}\,.
\end{equation}
For normalized single dimple volume, we set $n_D = 1$.

\subsection{Dimple Surface Area}

Similar to the volume of the dimples, we calculate the surface area of a single dimple and normalize the total surface area of the dimples by the surface area of the particle. The surface area of a single dimple ${s}_d$ is calculated as the surface area of a spherical cap formed between the central and valence spheres. For spheres of radii $R_+, R_-$, the surface area is given by
\begin{equation}
{s}_d = \tfrac{\pi R_{-}}{d} (R_{+} - R_{-} + d) (R_{+} + R_{-} - d) \,.
\end{equation}
For $R_{+} = R_{-} = 1$,
\begin{equation}
{s}_d = \pi (2-d) \,.
\end{equation}
The surface area of the particle, ${s}_{sp}$, is given by
\begin{equation}
{s}_{s\!p} = {s}_{P\!S} - (n_D \times {s}_{d}) \,.
\end{equation}
where ${s}_{P\!S}$ is the surface area of the original central sphere. This gives
\begin{equation}
{s}_{s\!p} = 4 \pi R_+^2 - n_D \times\pi (2-d) \,.
\end{equation}

The self-assembled non-spherical crystal structure is rather dependent on the number and arrangement of dimples in the particle.
We normalize the total dimple surface area ${S}_{d}$ for a particle as 
\begin{equation}
{S}_{d} = \frac{n_D \times {s}_{d} }{{s}_{s\!p}} \,.
\end{equation}

\subsection{Dimple Circumference}

Similarly, the circumference of the dimples ${c}_d$ is calculated from the
radius of the circle of intersection between the central and valence spheres.
For spheres of radii $R_+,R_-$, the circumference is given by
\begin{multline}
{c}_d = \tfrac{\pi}{d} \sqrt{(-R_{+} + R_{-} + d)(R_{+} - R_{-} + d)}\times\\ 
\sqrt{(R_{+} + R_{+} - d)(R_{+} + R_{-} + d)}
\end{multline} \,.
For $R_{+} = R_{-} = 1$,
\begin{equation}
{c}_d = \pi\sqrt{4-d^2} \,.
\end{equation}
This circumference is normalized by the distance between the central and valence
spheres. To reduce the numerical ratio, we use a weight $2\!\pi$ and call this
normalization factor ${c}_{s\!p}$. This factor is given by
\begin{equation}
{c}_{s\!p} = 2 \pi d \,.
\end{equation}

The total circumference of the dimples $C_d$ is thus normalized by the
circumference of a circle describing the distance between the parent and valence
sphere in the particle.
\begin{equation}
{C}_{d} =  \frac{n_D \times {c}_{d} }{{c}_{s\!p}} \,.
\end{equation}

\section{Critical Dimpling Amount}
\label{Appendix:Criticality}
\vskip-6pt
\begin{table}[ht] 
\caption{Crossover Values for Putative Densest Packings and Thermodynamic Assemblies} 
\centering 
\begin{tabular}{c c c c c} 
\multicolumn{5}{c}{ } \\
\hline\hline 
Particle & $f_{D\!P}^*$  & $C_d$ & $f_{S\!A}^*$ & $v_d$ \\ [0.5ex] 
\hline \hline 
P2 & $0.73$ & $1.21$ & $0.42$ & $0.0742$  \\ 
P3 & $0.53$ & $1.22$ & $0.48$ & $0.0714$ \\ 
P4t & $0.35$ & $1.20$ & $0.53$ & $0.0782$ \\ 
P4s & $0.66$ & $1.32$ & $0.73$ & $0.0769$ \\ 
P6 & $0.30$ & $1.26$ & $0.70$ & $0.0783$ \\ [0.5ex] 
\hline 
\end{tabular} 
\label{table:criticality_DP_SA} 
\vskip-12pt
\end{table}

The critical dimpling amount for the putative densest packing $f_{D\!P}^*$ and
self-assembly $f_{S\!A}^*$ of each type of multi-dimpled particle is the
crossover point in the corresponding particle behavior. At these crossover
values, different geometric characteristics of each particle type have been
computed as shown in Table \ref{table:criticality_DP_SA}.

To understand the packing of multi-dimpled spheroidal particles, we determine
the primary feature in the shape of these particles that affects their behavior.
We investigate the value of $f_{D\!P}^*$ compared to the total volume ${V}_{d}$,
surface area ${S}_{d}$ and circumference ${C}_{d}$ of all dimples on a single
particle as $d$ is varied.  We normalize each of these three parameters. The
volume and surface area are normalized by the remaining respective quantity in a
single particle, while the circumference is normalized by the circumference of a
circle with radius equal to the distance between the central and valence
spheres, a linear function of the depth of the dimple. We compute the total
normalized dimple volume, surface area and circumference to depth ratio in each
family of particle at $f_{D\!P}^*$. We find that the total circumference to
depth ratio ${C}_{d}$ of the dimples is $1.23\!\pm\!0.06$ at $f_{D\!P}^*$
across all five particle families; see Fig.\ \ref{fig:scaling_DP}. In contrast, no
dimple volume or surface area showed no similar collapse.

We also observed that in self-assembly, all particle types undergo a structural
transition from BCC in which particles have nearly isotropic pairwise
interactions, into a phases where particle symmetry is important.
To quantify when the concave features are sufficient to induce a
change in the preferred structure, we compare particle shape characteristics at
the critical dimpling amount for self assembly $f_{S\!A}^*$. We find that,
across particle types, $v_d$ shows little variation at $f_{S\!A}^*$, which
suggests that \emph{the volume of individual dimples controls assembly behavior,
independent of particle type.} 
\vskip-6pt
\begin{table}[ht] 
\caption{Geometric and Numerical Calculations of Particle Interlocking} 
\centering 
\begin{tabular}{c c c c c} 
\multicolumn{4}{c}{ } \\
\hline\hline 
Particle & Description & $f^*_{D\!P}$ & $f^\dagger_{D\!P}$  & $d^2$ \\ [0.5ex] 
\hline \hline 
P2 & Bivalent & $0.73$ & $0.86$ & $2.0$ \\ 
P3 & Trivalent & $0.53$ & $0.70$ & $2.14$ \\ 
\hline 
\end{tabular} 
\label{table:encompass} 
\vskip-12pt
\end{table} 

\section{Analytic Calculations of Packings}
\label{analyticPackings}

\begin{figure}
  \includegraphics[width=\columnwidth]{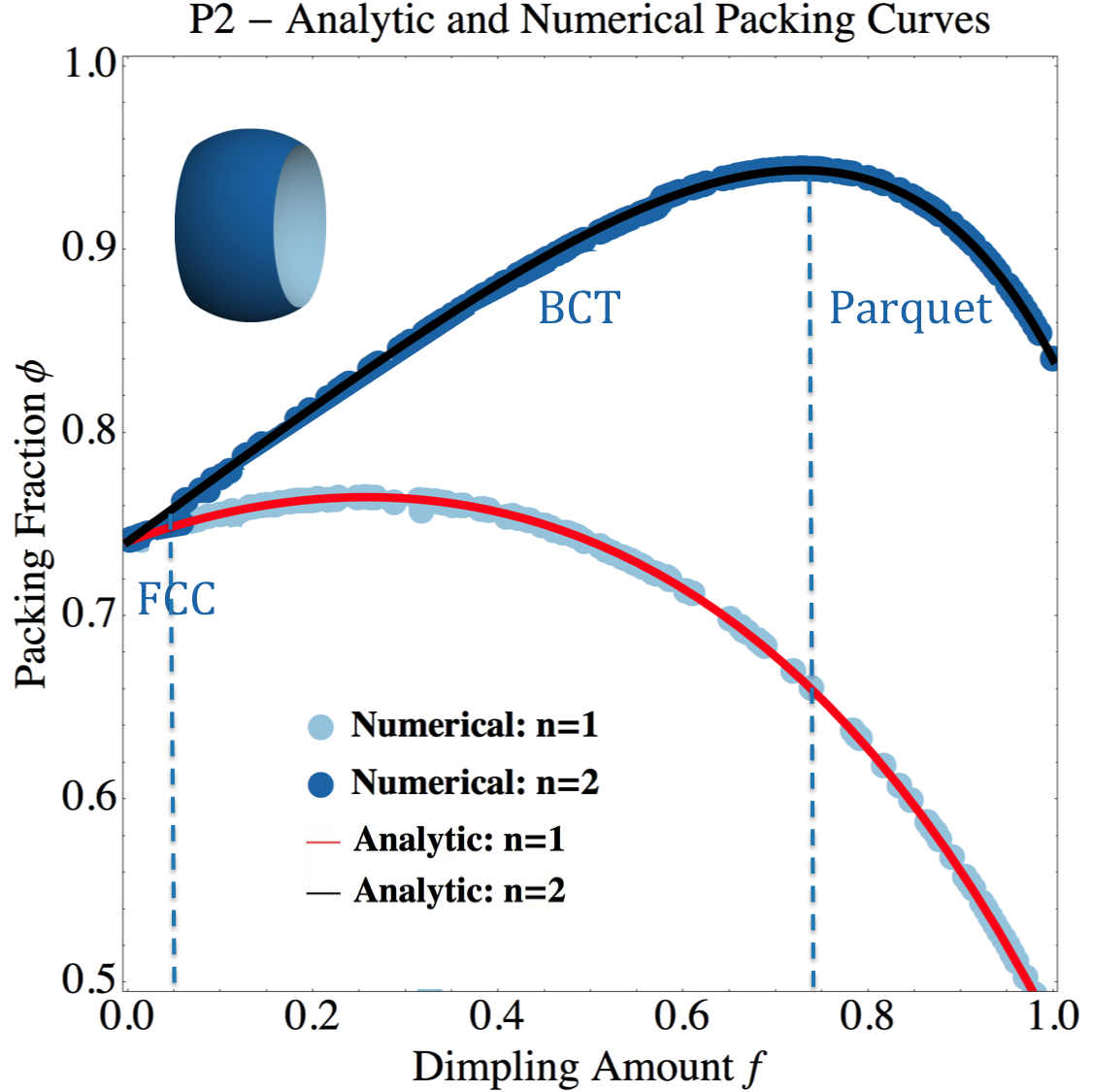}
  \caption{Analytic and numerical packing curves showing a perfect fit for bivalent particle (P2) for one and two particles in a box. Analytic packing curve for one particle is shown in red, two particles in black. The dark markers denote numerical densest packing calculations for two particles in a box, and light for one particle. Blue dashed lines show the different packing regions found in numerical calculations.}
  \label{fig:P2N_compare}
\end{figure}

\begin{figure}
  \includegraphics[width=\columnwidth]{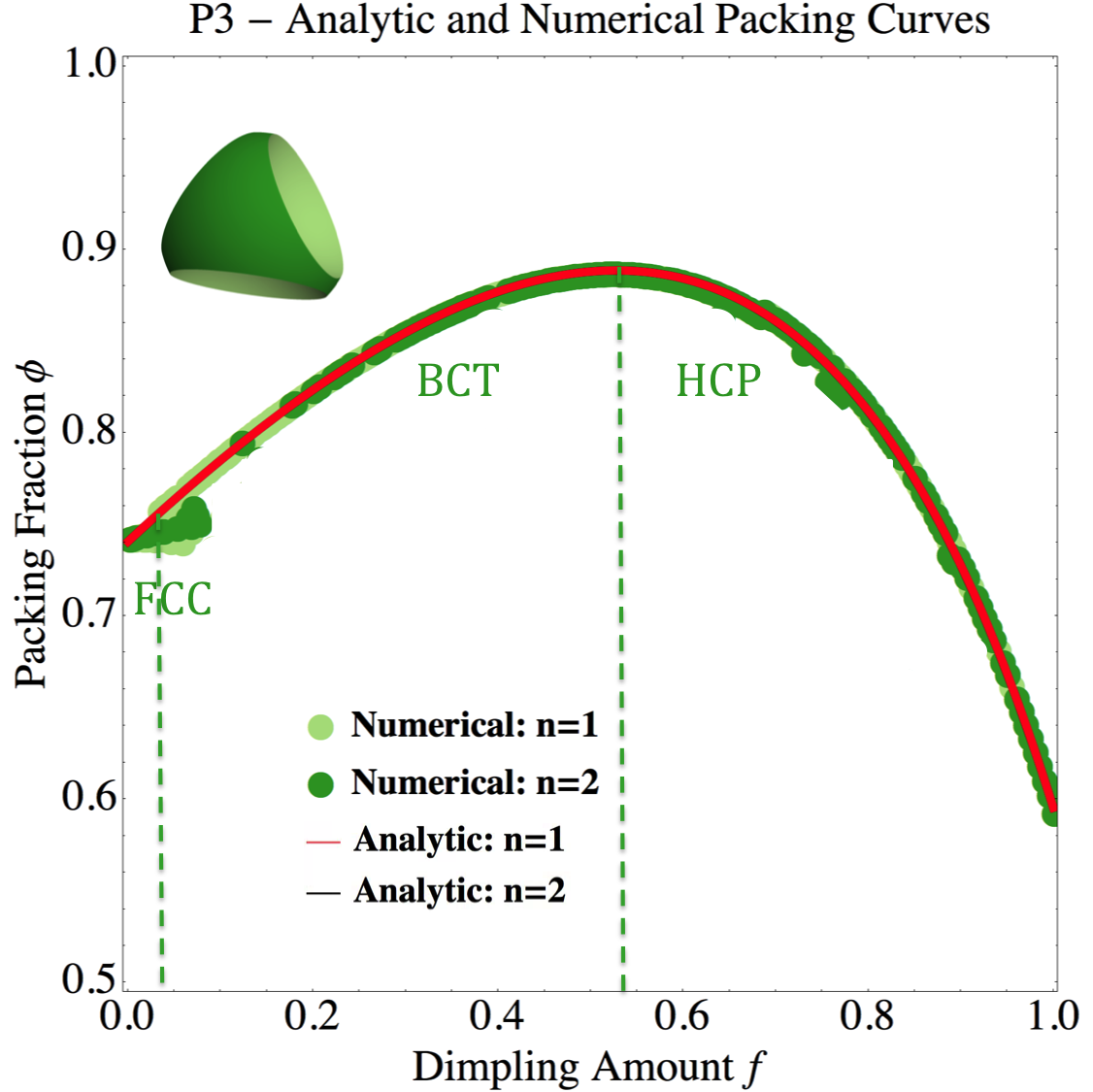}
  \caption{Analytic and numerical packing curves showing a perfect fit for trivalent particle (P3) for one and two particles in a box. The curves overlap as one and two particles pack in the same manner. Green dashed lines show the different packing regions found in numerical calculations.}
  \label{fig:P3N_compare}
\end{figure}

\begin{figure}
  \includegraphics[width=\columnwidth]{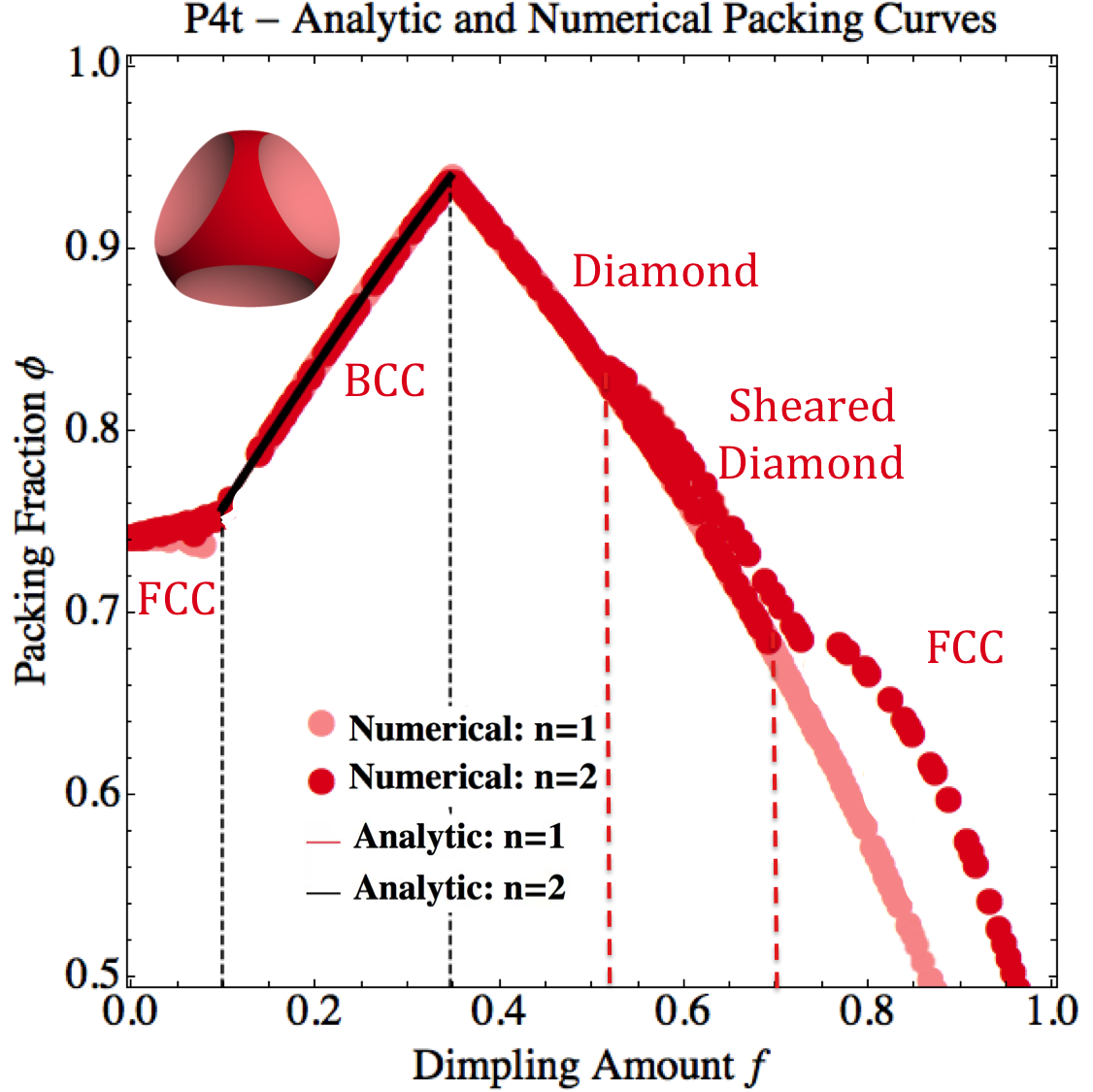}
  \caption{Analytic and numerical packing curves showing a perfect fit in a single region between black dashed lines for tetrahedral tetravalent particle (P4t) of one and two particles in a box. Analytic packing equations could not be calculated at all dimpling amounts by analytic means. Red dashed lines show the different packing regions found in numerical calculations, in addition to those found in analytic calculations.}
  \label{fig:P4Nth_compare}
\end{figure}

\begin{figure}
  \includegraphics[width=\columnwidth]{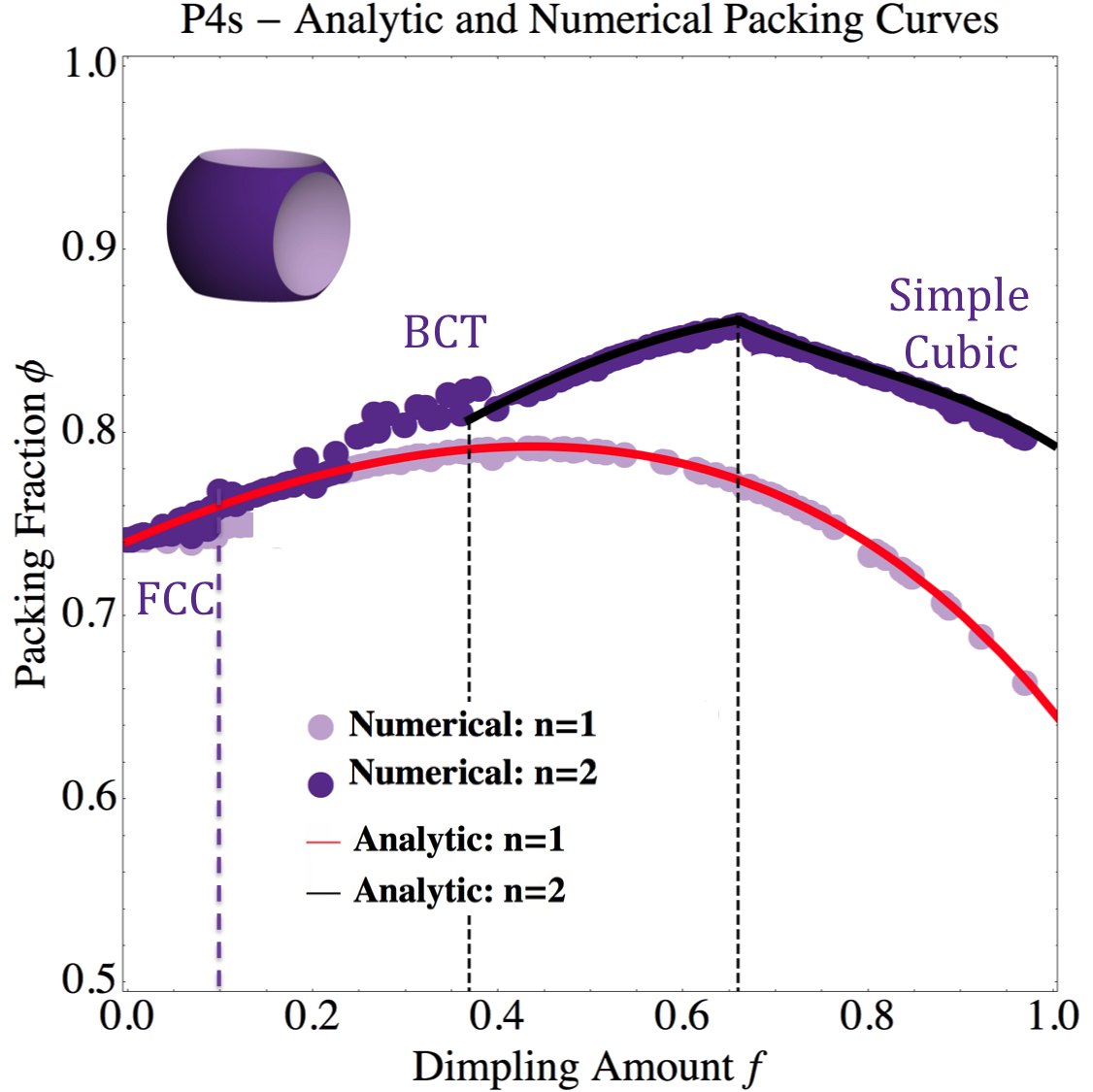}
  \caption{Analytic and numerical packing curves of a square tetravalent particle (P4s) showing a perfect fit in the entire domain for one particle in a box and between black dashed lines in two regions for two particles in a box. Analytic packing curves could not be calculated at all dimpling amounts for two particles by analytic means. Magenta dashed lines show the different packing regions found in numerical calculations, in addition to those found in analytic calculations.}
  \label{fig:P4Ns_compare}
\end{figure}

\begin{figure}
  \includegraphics[width=\columnwidth]{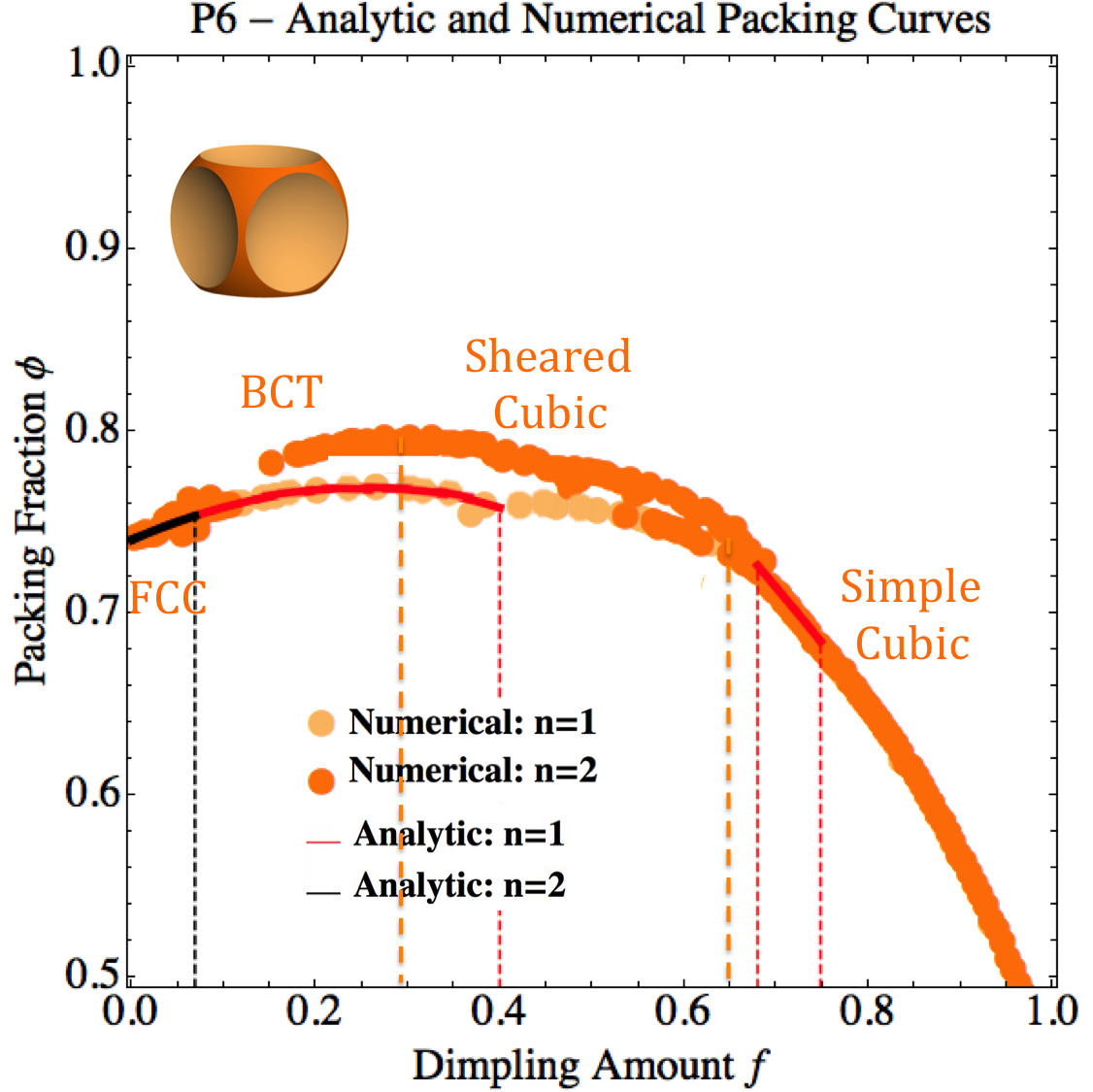}
  \caption{Analytic and numerical packing curves of a hexavalent particle (P6) showing a perfect fit between red dashed lines in two regions for one particle in a box and between black dashed lines in one region for two particles in a box. Analytic packing curves could not be calculated at all dimpling amounts for two particles by analytic means. Orange dashed lines show the different packing regions found in numerical calculations, in addition to those found in analytic calculations.}
  \label{fig:P6N_compare}
\end{figure}

\begin{figure*}
\includegraphics[scale=0.250]{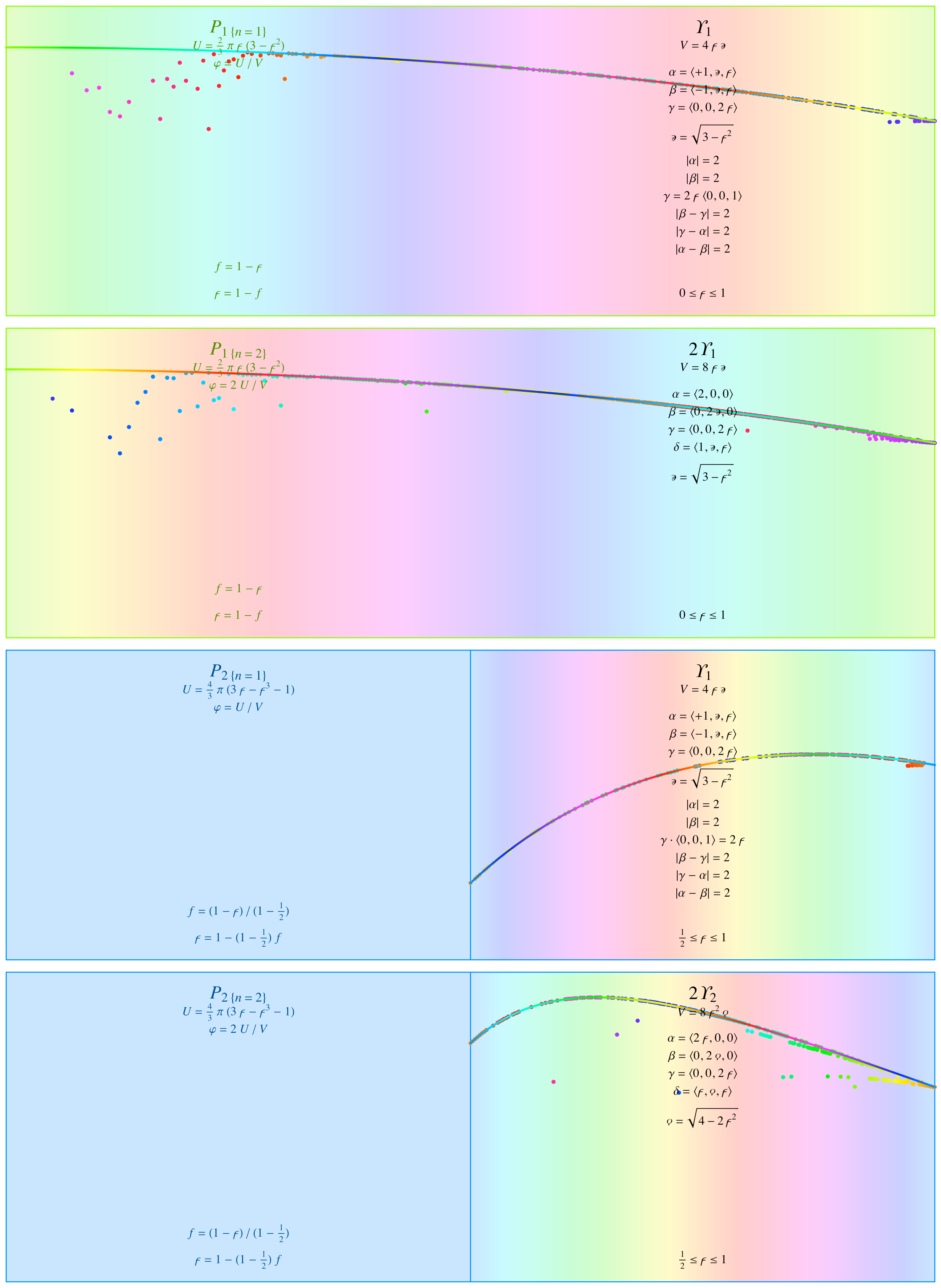}
\caption{Analytic packing density ${3\over 10}\le\phi\le 1$ and packing curves as a function of thickness parameter $0\le\digamma\le 1$ ({\it right})
for single \& double lattice packings of particles P1 \& P2.}
\label{fig:P1P2_analytic}
\end{figure*}

To verify the putative densest packing predicted computationally via MC
simulations for the particles studied, analytic calculations of the packings of
these particles were performed.  In Figures \ref{fig:P2N_compare} -
\ref{fig:P6N_compare}, we plot analytic calculations of the packing curve and
numerically predicted packings for one and two particles of all particle types.
We show that our numerical calculations match the analytic findings to within
$0.001\%$ when $f \geq 0.10$.  Additionally, we perform
numerical and analytic packing calculations for single dimpled particles (see
Figure \ref{fig:P1P2_analytic}).  We find that single dimpled particles do not
follow the pattern shown by other particles in this study due to the lack of
discrete symmetry in their shape.

\begin{figure*}
\includegraphics[scale=0.250]{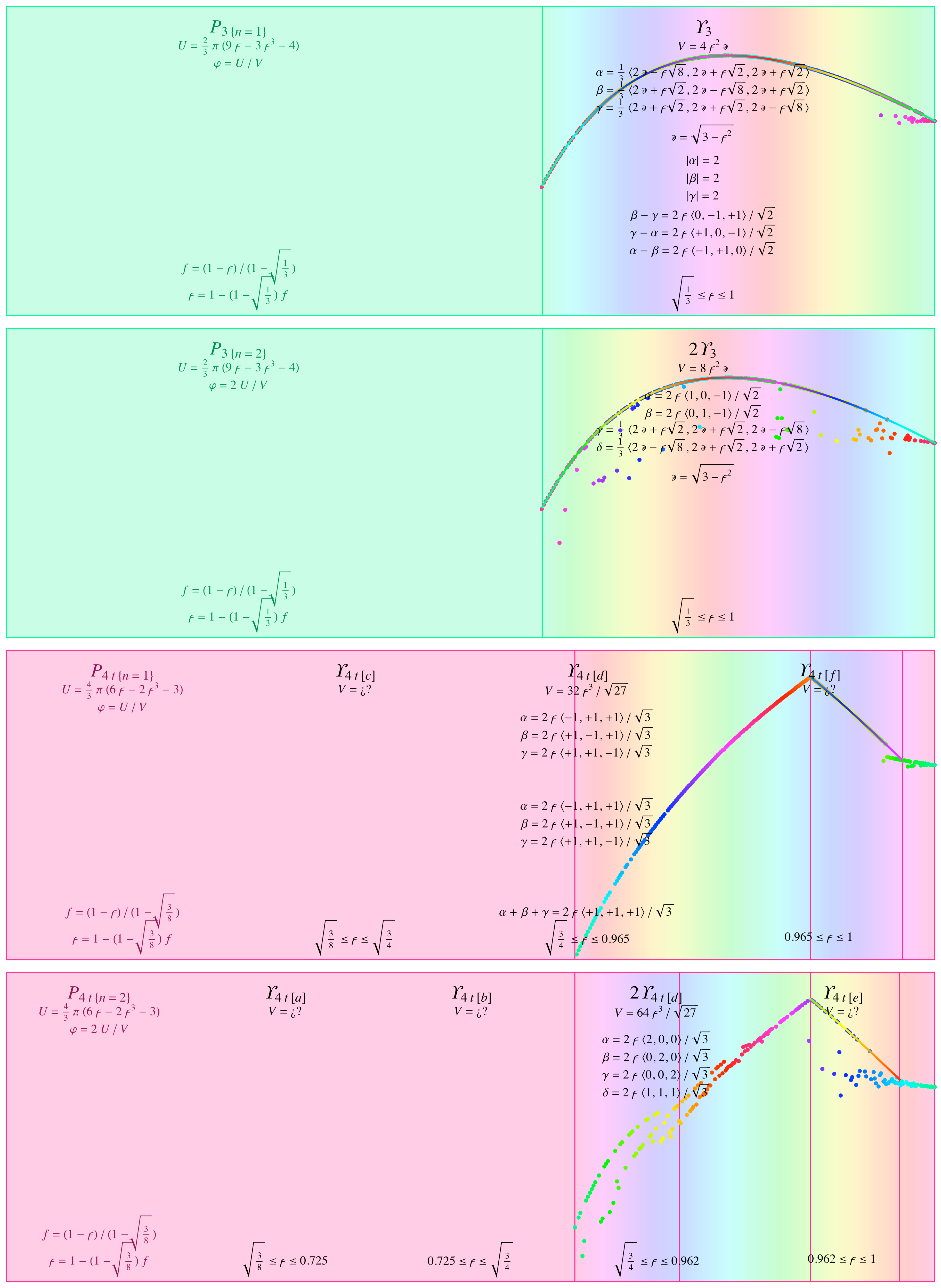}
\caption{Analytic packing density ${3\over 10}\le\phi\le 1$ and packing curves as a function of thickness parameter $0\le\digamma\le 1$ ({\it right})
for single \& double lattice packings of particles P3 \& P4t.}
\label{fig:P3P4t_analytic}
\end{figure*}

\begin{figure*}
\includegraphics[scale=0.250]{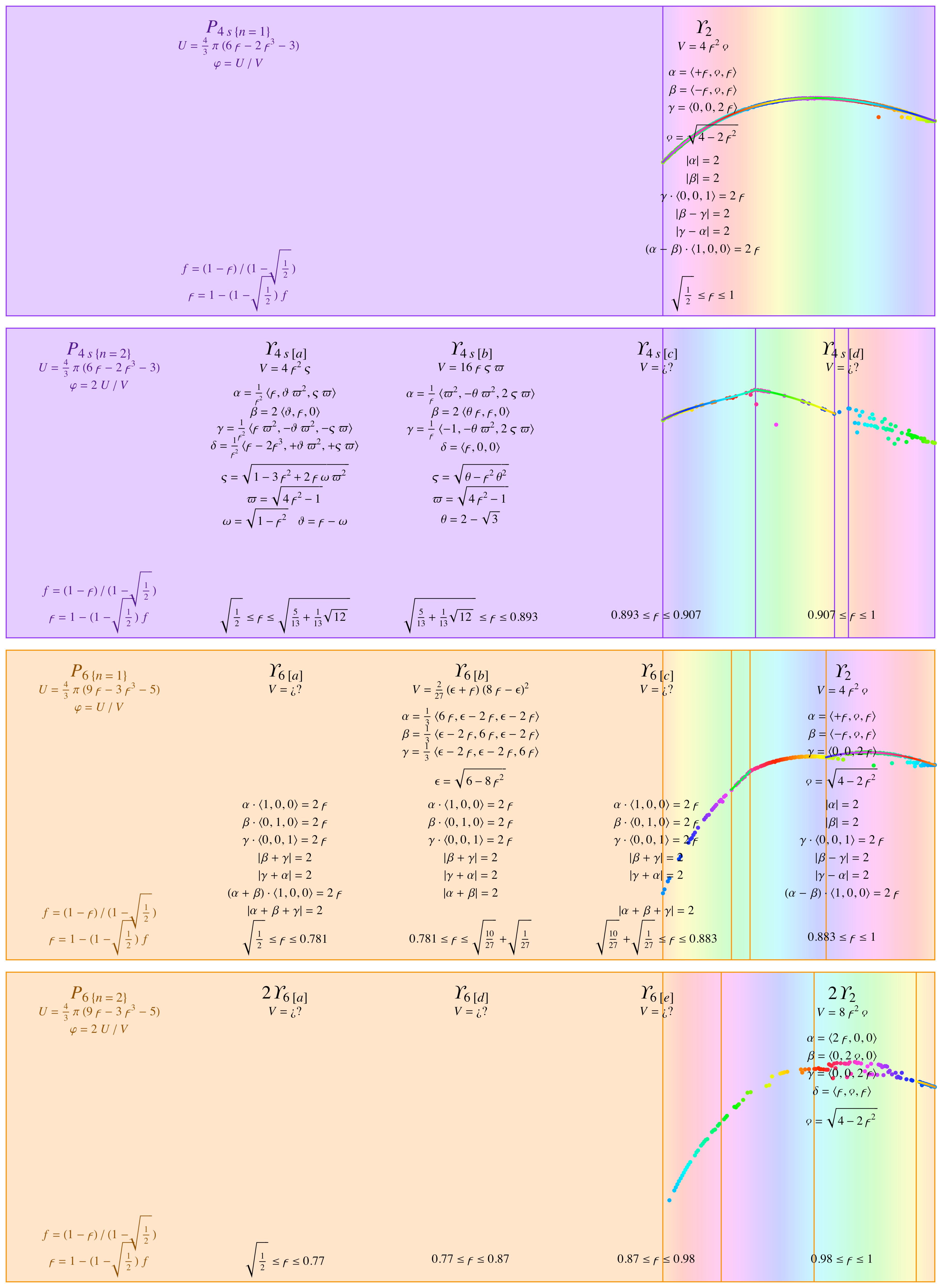}
\caption{Analytic packing density ${3\over 10}\le\phi\le 1$ and packing curves as a function of thickness parameter $0\le\digamma\le 1$ ({\it right})
for single \& double lattice packings of particles P4s \& P6.}
\label{fig:P4sP6_analytic}
\end{figure*}

We further numerically calculate maximum packing densities of all particle types
with up to eight particles.  However, due to the complexity of the analytic
packing problem for number of particles $n \geq 3$, we do analytic calculations
only for (some) one- and two-particle packings ($n = 1,2$) for different
particle types. If solvable, these analytic curves and intersection equations
are shown in Figures \ref{fig:P1P2_analytic} - \ref{fig:P4sP6_analytic}.  For
these results it is useful to work in terms of a linearly shifted parameter
$\digamma = 0.5d$, where $d$ is the distance parameter, rather than the dimpling
parameter $f$ we use in the main text. We denote the volume of a particle by
$U$, the volume of a unit cell of the lattice by $V$, the packing fraction by
$\varphi$, and different packing regions by $\Upsilon$. We denote the vectors
describing a unit cell of the packing as $\alpha$, $\beta$, and $\gamma$, and
the volume is given by the standard formula $V=\alpha\cdot(\beta\times\gamma)$.
Images are colored so that the color on the left hand side of the image matches
the color used for the particle in the rest of the text, and then proceeds
through the visual spectrum over the range of geometrically allowed particle
parameters starting toward the red end of the spectrum for single particle
packings, and the blue end of the spectrum for two particle packings.


The relationship between the thickness parameter $\digamma$ and dimpling
parameter $f$ for each particle type is included in the bottom left side of the
graph. Analytic calculations of the dimensions of the packing box are shown in
the graph.  The analytic packing density curve is drawn as a line in these
figures against the thickness parameter $\digamma$, in regions where it can be
calculated. In order to compare with putative density calculations, these lines
are superimposed on markers denoting numerical calculations. Various regions in
these analytic packing curves are explicitly marked through vertical lines at
different values of $\digamma$, the thickness parameter. In regions where the
analytic curves can be calculated, we find that they match our numerical
calculations, with an error less than $0.001\%$ when $f \geq 0.10$. We include
the single dimpled particle P1, since the same lattice $\varUpsilon_1^{}$ is
shared between particles P1 and P2. Similarly, the same lattice
$\varUpsilon_2^{}$ is shared among particles P2, P4s and P6. 

\section{Self-Assembly}
\label{Appendix:SATrans}

\subsection{Transition Density}

\begin{figure}
  \includegraphics[width=0.95\columnwidth]{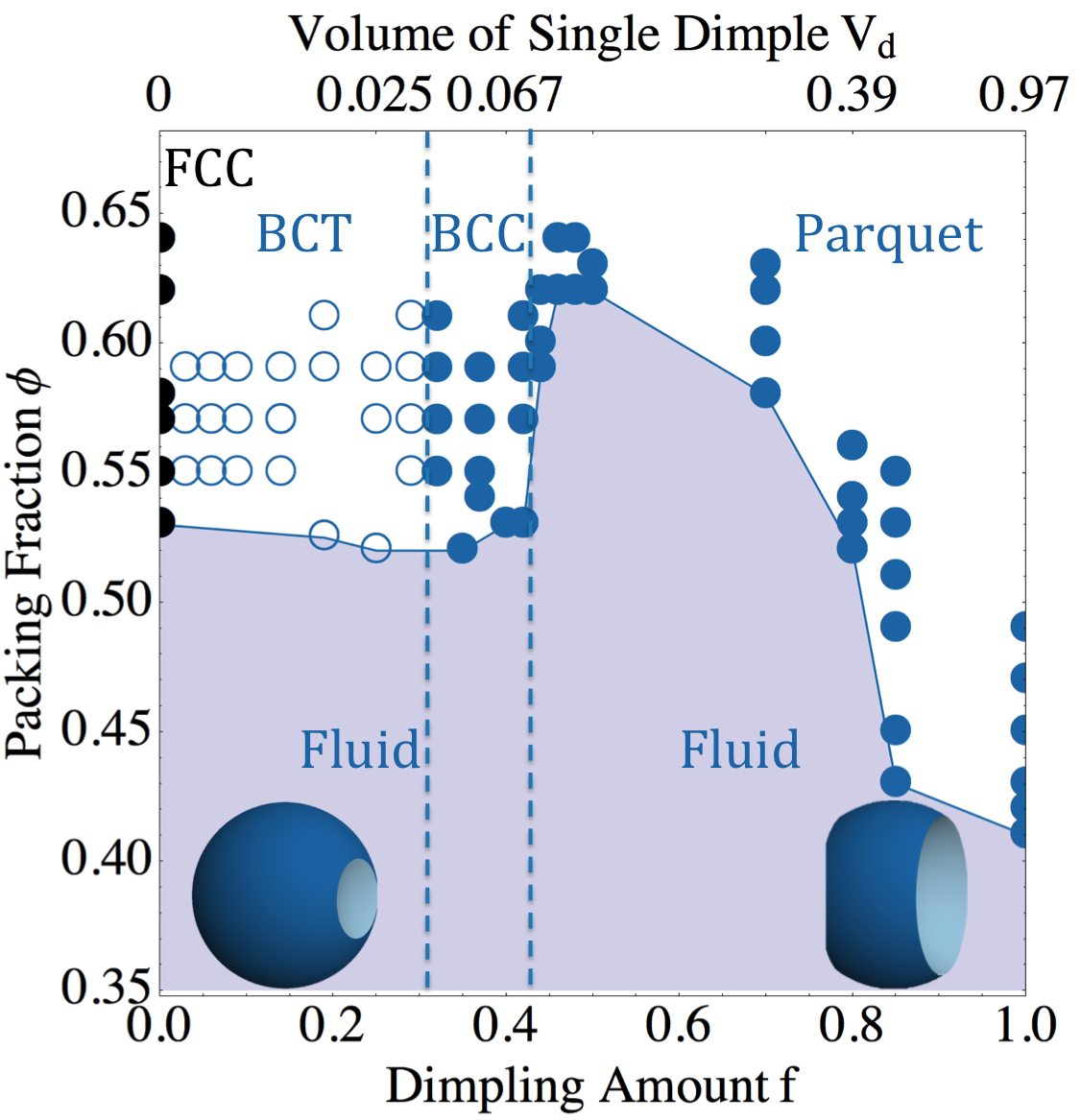}
  \caption{Minimum packing fraction at which bivalent particles (P2) assemble crystal structures. The structures assembled vary with dimpling amount $f$, from FCC for hard sphere to BCT, BCC and a parquet structure. Blue dashed lines show the different structural regions found in our simulations.
  \label{fig:P2N_TransD}}
\end{figure}

\begin{figure}
  \includegraphics[width=0.95\columnwidth]{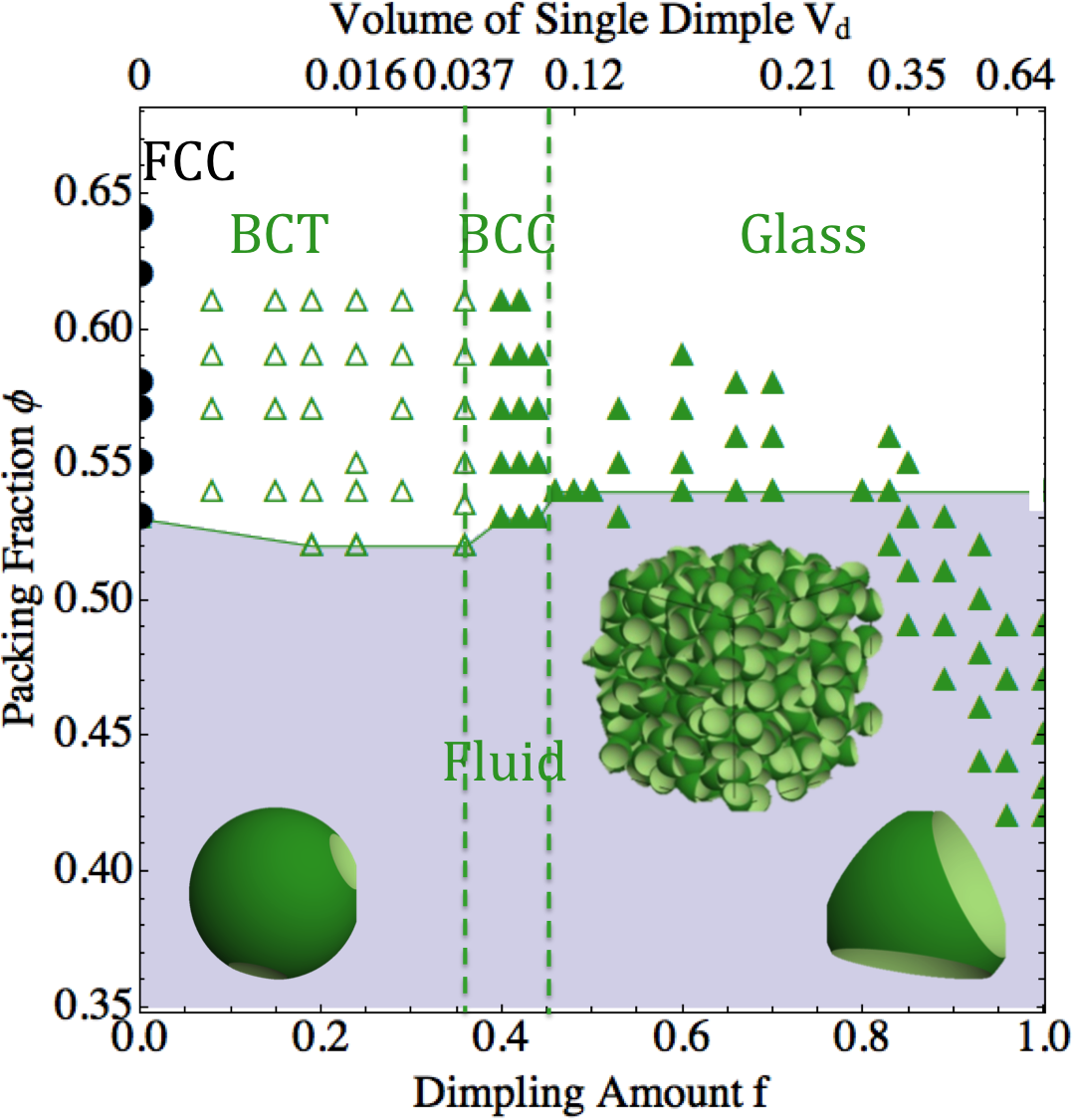}
  \caption{Minimum packing fraction at which trivalent particles (P3) assemble crystal structures. The structures assembled vary with dimpling amount $f$, from FCC for hard sphere to BCT, BCC and then form glass. Green dashed lines show the different structural regions found in our simulations.
  \label{fig:P3N_TransD}}
\end{figure}

\begin{figure}
  \includegraphics[width=0.95\columnwidth]{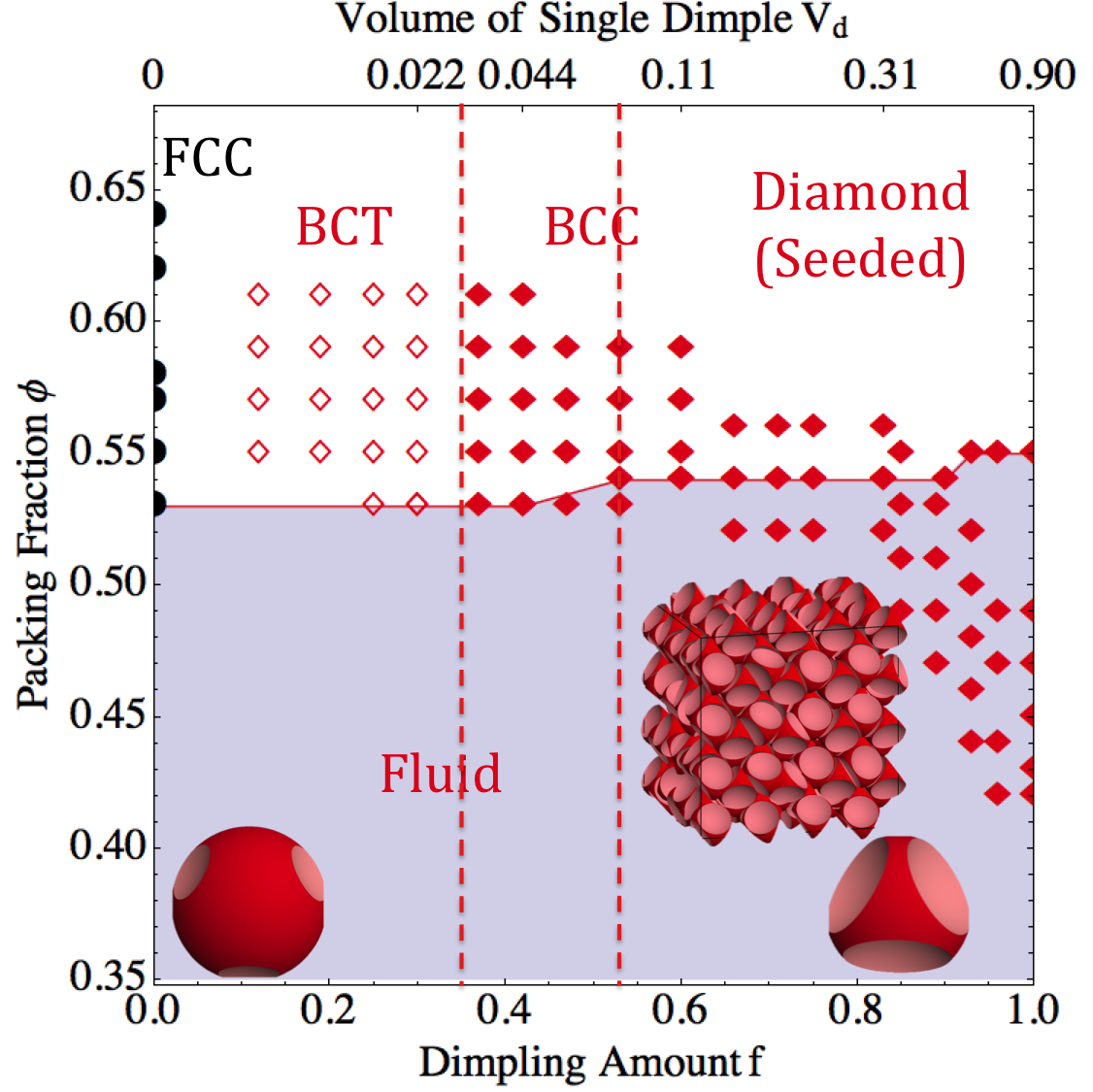}
  \caption{Minimum packing fraction at which tetrahedral tetravalent particles (P4t) assemble crystal structures. The structures assembled vary with dimpling amount $f$, from FCC for hard sphere to BCT, BCC and diamond. Red dashed lines show the different structural regions found in our simulations.  \label{fig:P4Nth_TransD}}
\end{figure}

\begin{figure}
  \includegraphics[width=0.95\columnwidth]{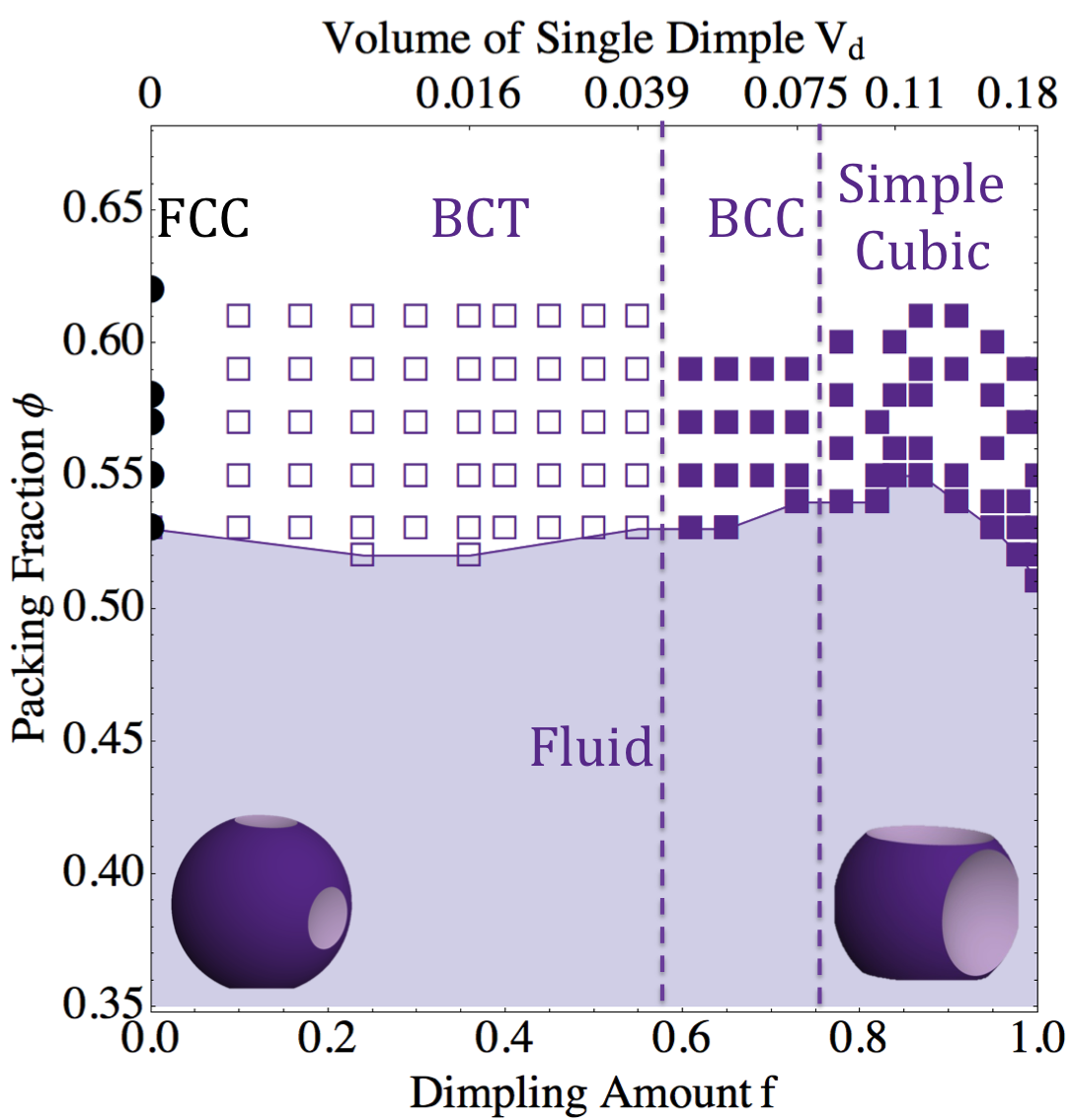}
  \caption{Minimum packing fraction at which square tetravalent particles (P4s) assemble crystal structures. The structures assembled vary with dimpling amount $f$, from FCC for hard sphere to BCT, BCC and simple cubic. Purple dashed lines show the different structural regions found in our simulations.
  \label{fig:P4Ns_TransD}}
\end{figure}

\begin{figure}
  \includegraphics[width=0.95\columnwidth]{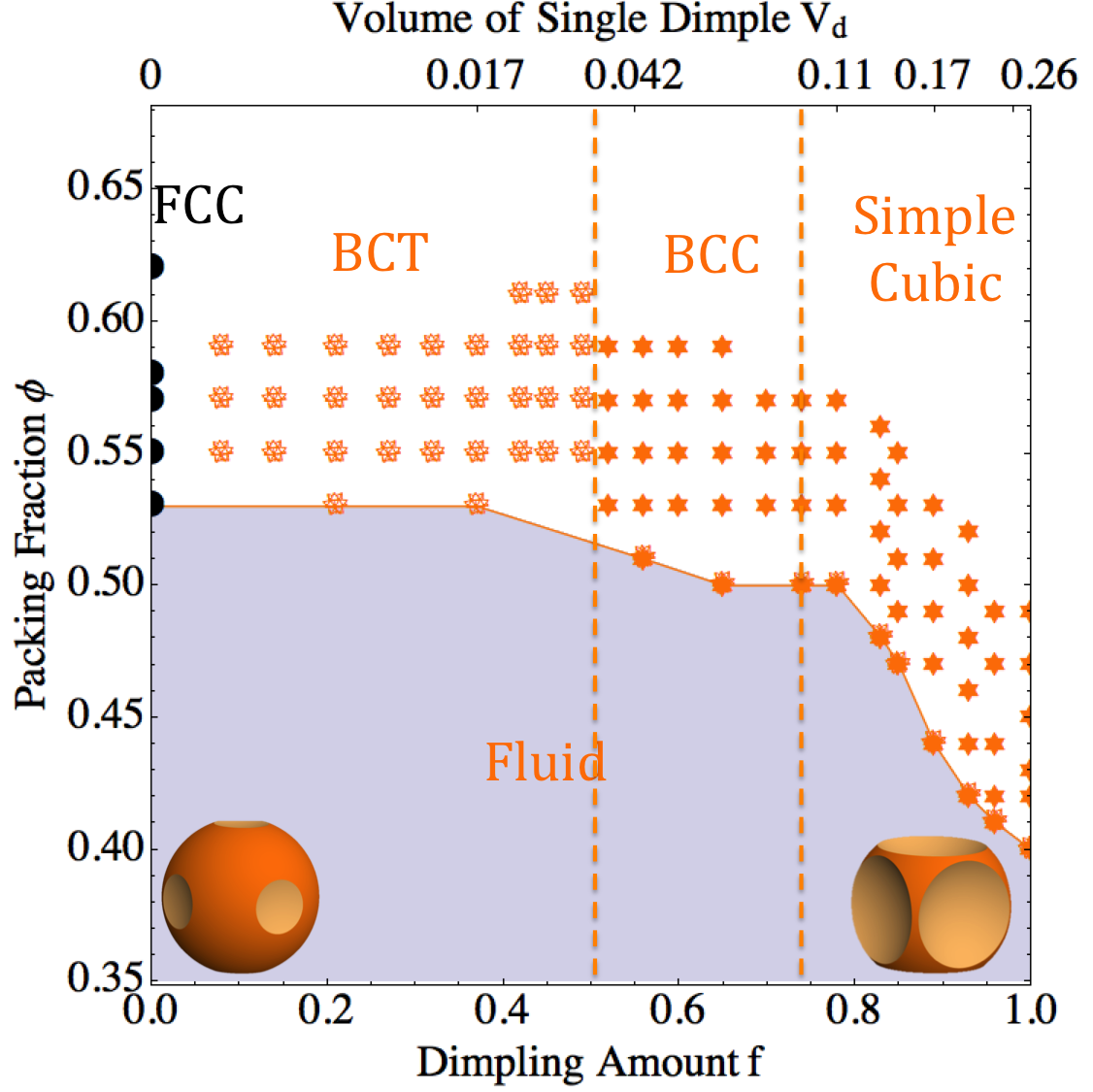}
  \caption{Minimum packing fraction at which hexavalent particles (P6) assemble crystal structures. The structures assembled vary with dimpling amount $f$, from FCC for hard sphere to BCT, BCC and simple cubic. Yellow dashed lines show the different structural regions found in our simulations.
  \label{fig:P6N_TransD}}
\end{figure}

The particles studied in this work assemble at approximately the same
conditions. For the sake of completeness, we show the transition density phase
diagrams for thermodynamic self-assembly of each particle in Figures
\ref{fig:P2N_TransD} -- \ref{fig:P6N_TransD}. In these plots, we also observe
the reported behavior of a crossover in the assembly of particles beyond a
constant critical volume of a single dimple in the top axis.
 
These crystal structures are formed by entropic interactions between the convex
and concave regions of two adjacent particles.  We find that the emergent
valence is dependent on the geometry of the dimples (see section
\ref{Appendix:PMFT} for details).  The number of directions for the alignment of
the particles depends on the crystal lattice, for example, two in the parquet
structure.  It should also be noted that the particles have rotational
symmetries, which align with the symmetry of the crystals and thus do not break
the symmetry of the crystals. In the case of the tetrahedrally tetravalent
particle, particles have two orientations in the diamond crystal structure
formed.

The self-assembly of multi-dimpled spheres into crystal structures occurs at
packing fractions of $40 - 50\%$, much lower than that observed in entropically
driven assemblies of convex polyhedra, typically $> 50\%$. These assemblies are
thus candidate structures for entropically assembled porous structures, and can
find applications such as advanced catalysis. By designing porous structures
using entropy alone, their assembly pathways have lesser kinetic traps than
their enthalpic counterparts, giving rise to structures that can be easily
assembled and reconfigured.


\subsection{Dimpling Amount and Volume of a Dimple}

As shown in Figures \ref{fig:P2N_TransD} -- \ref{fig:P6N_TransD}, the phase diagrams do not collapse at a constant dimpling amount $f$. However the volume of a single dimple is constant at the crossover across all phase diagrams. 
Here we include calculations that show the relationship between volume of a single dimple and the dimpling amount, in order to show that these two quantities are not linear functions of each other.

For bivalent particle, P2, with two dimples, we have:
\begin{equation}
  \begin{split}
    f &= \frac{4-d^2}{4-1} \; , \\
    d^2 &= 4 -3f \; .
  \end{split}
\end{equation}

where, $f$ is a linear mapping of $d^2$, the distance between the valence and central spheres.

Next, we have:
\begin{equation}
  \begin{split}
    V_d &= \frac{v_d}{v_{sp}} \; , \\
    V_d &= \frac{(d+4)(2-d)^2}{8-(d+4)(2-d)^2} \; .
  \end{split}
\end{equation}
where, $V_d$ is the normalized single dimple volume, equivalent to the ratio of the volume of a single dimple to the volume of the remaining particle.

In a similar fashion, we understand that the relationship between the volume of a single dimple and the dimpling amount is different for each particle type and is a function of the dimple geometry in a particle.

\bibliographystyle{apsrev4-1}
\bibliography{Sphinx.bib}

\end{document}